\documentclass[%
reprint,onecolumn,
 amsmath,amssymb,
 aps,
]{revtex4-2}

\usepackage{graphicx}
\usepackage{subcaption}
\usepackage{dcolumn}
\usepackage{bm}
\usepackage{empheq}
\usepackage{natbib}
\usepackage{xspace}
\usepackage{verbatim}
\usepackage{graphicx}
\usepackage{array}
\usepackage[space]{grffile}
\usepackage{latexsym}
\usepackage{amsfonts,amsmath,amssymb}
\usepackage{url}
\usepackage[utf8]{inputenc}
\usepackage[normalem]{ulem}
\usepackage{float} 
\usepackage{siunitx}
\usepackage{hyperref}
\usepackage{adjustbox} 
\usepackage{orcidlink} 

\usepackage{filecontents}
\usepackage{appendix}
\usepackage{xcolor}
\newcommand\apjl{{\@eapj@ApJLetters}}

\newcommand{\duvcom}[1]{{\textcolor{red}{#1}}}

\newcommand{\oth}[1]{\overset{\tiny{\circ}}{#1}}

\begin{document}
\title{Probing ALP-photon couplings in Neutron Stars:\\  scalar versus pseudoscalar cases}
            
\author{D. Suárez-Fontanella}
 \email{duvier@usal.es}
\author{C. Albertus}
 \email{albertus@usal.es}

\author{M. \'Angeles P\'erez-Garc\'ia }
 \email{mperezga@usal.es}
\affiliation{Department of Fundamental Physics and IUFFyM, University of Salamanca, 
Plaza de la Merced s/n, E-37008 Salamanca, Spain.}

\date{\today}

\begin{abstract}
We investigate the distribution of an interacting axion-like massive field within a magnetized Neutron Star. For this we consider the effect of an intense density-dependent axially symmetric stellar magnetic field ${\bf B}(r,\theta)$ adding another much weaker, but non-vanishing,  electric field.  We particularize the latter for the case when a finite chiral charge density is present. The axion field is thus coupled to a generic function $Q(F_0,G_0)$ depending on Lorentz invariants $F_0, G_0$ which can be constructed from these electromagnetic fields. From this, the static axion field equations are solved as function of stellar radial coordinate and angular direction,  $a(r,\theta)$, using a prescribed linear form for $Q$. In addition, we use a semi-analytical approach to calculate the stellar structure in this hybrid system where pressure components are treated under a perturbative scheme, provided induced deformations with respect to spherical symmetry are tiny. Our results show that the axion couplings to magnetic and  electric fields along with its mass, critically determine the axion spatial distribution. Furthermore, we  focus on the possibility that the axion field might accumulate in specific outer regions of the star, particularly within the crust, where it could form a condensate. We explore the possible presence of  magnetic flux tubes from superconductor phases in this outer layers and qualitatively show they may enhance local conversion into photons. We explore prospects of detectability through indirect methods. 

\end{abstract}

\maketitle

\section{Introduction}\label{Intro}

Neutron stars (NS) serve as exceptional natural laboratories for studying physical phenomena under extreme conditions. With densities above that of atomic nuclei and magnetic fields reaching intensities of $\sim 10^{15}$ G in the surface of magnetars to $\sim 10^{18}$ G in the interior \cite{maperezPhysRevC.77.065806,maperezPhysRevC.91.045803}, these objects provide a unique opportunity to explore fundamental interactions beyond the Standard Model (SM) of particle physics. In particular the problem of finding the dark matter (DM) candidate able to accommodate the mounting evidence for a dark sector continues. In this regard, among the proposed extensions to the SM, low mass candidates seem popular in the literature\cite{cirelli2024dark} i.e. axions, axion-like particles and other light particles, generically known as weakly interacting slim particle (WISPs) have garnered significant interest, see \cite{Carenza_2025} for a review. 
One of these light DM dark matter candidates with increasing popularity is the QCD axion, a particle introduced to resolve the strong CP problem through the Peccei-Quinn mechanism, for a recent review see  \cite{ringwald2024review}. In detail it consists of adding a new particle $a$ with the coupling

$$
\frac{a}{f_a} \frac{g_s^2}{32 \pi^2} G^{\mu \nu} \tilde{G}^{\mu \nu}
$$

where $g_s$ is the strong coupling constant, $G^{\mu \nu}$ is the gluon field strength and $\tilde{G}^{\mu \nu}=$ $\frac{1}{2} \epsilon^{\mu \nu \rho \sigma} G_{\rho \sigma}$. The above coupling dynamically sets the neutron electric dipole moment (EDM) to zero. At low energies, the axion obtains a potential, with a period of $2 \pi f_a$, from the coupling to gluons\cite{GrillidiCortona:2015jxo}.

These scalar particles have a mass in the range $\sim 10^{-11} \mathrm{eV}-1 \mathrm{eV}$ ($c=1$) for the QCD to be DM. Less restrictive than the QCD axion is  a class of so-called axion-like particles (ALPs) that have triggered a lot of interest both on the theory and on the experimental side \cite{ringwald2024review}. 

In this work we will focus on ALPs in the mass range $m_a\gtrsim 10^{-11}$ eV. Thus ALPs are expected to exhibit clustering in the late Universe, participating in the formation of the non-linear structures, including the dark matter halos \cite{Centers_2021}.  ALPs simulations seem indicate that their  density in the solar neighborhood in the Milky Way, compared to the average density of ALPs in the current Universe, does not differ significantly from that of cold DM.  Besides, most works on ALPs assume it is  comprised of a single species, being itself a significant portion of the dark matter. However, the assumption of a single field should be considered a toy model. The generic prediction from string theory is of an axiverse of ALPs \cite{Arvanitaki_2010}.

The axion-photon interaction is usually characterized by a coupling to a function $Q(F_0,G_0)$ depending on electromagnetic Lorentz invariants $F_0, G_0$ which can be constructed from stellar electric and magnetic fields ${\bf E}, {\bf B}$ \cite{Escobar_2014}.  When solving the axion field equations  $Q$ displays a particular form depending on their  assumed scalar/pseudoscalar nature \cite{Plakkot_2023}.  As examples of previous works regarding coupling of pseudoscalar axions to electromagnetic fields, see for example \cite{anzuini2023magnetic} and references therein. In principle, conventional axion models lead to CP-even (CPE) axion-photon interaction of the form $\mathcal{L}_{a \gamma}^{C \mathrm{PE}}= g_{a \gamma \gamma} a F_{\mu \nu} \widetilde{F}^{\mu \nu}$ where the axion-photon coupling $g_{a \gamma \gamma}$ dictates the probability of axion-photon conversion in the presence of electromagnetic fields, as is probed in several axion detection experiments \cite{Irastorza_2018}. $a$ is the axion field and $F_{\mu \nu}$ and its dual are constructed from electromagnetic fields as we will explain later. CP violating (CPV) axion interactions allow  an additional effective axion-photon term to the Lagrangian under the form $
{L}_{a \gamma}^{\mathrm{CPV}}=\frac{\bar{g}_{a \gamma \gamma}}{4} a F_{\mu \nu} F^{\mu \nu}$ which causes an interaction proportional to $\sim a\left( \textbf{E}^2-\textbf{B}^2\right)$ as opposed to the conventional CP-even coupling proportional to $\sim a(\bf{E \cdot B})$. This will be explained in detail later.

There are already plenty of works in the literature based on General Relativistic Magneto-HydroDynamics (MHD)  simulations  \cite{Font2008LRR....11....7F} where the dynamical entanglement of electromagnetic fields in the interior of Neutron Stars allow the description of these highly complex system using multipole expansions  \cite{haskel10.1111/j.1365-2966.2008.12861.x, Sur_2020}. General magnetic field solutions usually include several components, typically, a poloidal one 
 $\boldsymbol{B}_{\mathrm{p}}=\left(B_r, B_\theta, 0\right)$ and a toroidal one $\boldsymbol{B}_{\mathrm{t}}=\left(0,0, B_\phi\right)$ in spherical coordinates.
The presence of an electric field inside the core of a NS has usually been neglected under the assumption of perfect conductor. Note that this is related, in practice, to the assumed infinite conductivity, which has been however proved not realistic as conductivities from protons and electrons are quite different and there is a gradient in composition \cite{Reise1995ApJ...442..749R}. 
As it turns out the  microscopic interior is key to understand the large-scale magnetic fields that, nevertheless remain poorly unknown. Even small deviations from the chemical equilibrium after the initial stellar deleptonization stage can cause additional contributions to  microscopic seeds of electromagnetic fields \cite{Palenzuela_2013} .

Actually, under slightly off-equilibrium conditions, the Adler-Bell-Jackiw (ABJ) chiral anomaly~\cite{adlerPhysRev.177.2426,Bell1969NCimA..60...47B} modifies the electromagnetic transport through an anomalous current contribution. This quantum effect generates an electric field anti-parallel to the magnetic field and proportional to the, presumably small, chiral asymmetry. This is known as the Chiral Magnetic Effect (CME)~\cite{dvornikov2019permanent,acosta2015chiral,rogachevskii2017laminar,dehman2024origin}. As we will explain later, a chiral chemical potential $\mu_5 \equiv \frac{\mu_L - \mu_R}{2}$ depending on different lepton fractions with left (L) and right (R) chirality drives a current density $\mathbf{J}_{\rm{CME}} = \frac{e^2}{4\pi^2} \mu_5 \mathbf{B}.$
The CME reflects a dynamic helicity transfer between fermionic and electromagnetic sectors, the latter encoding the topological structure of $\mathbf{B}$-field lines via the vector potential $\mathbf{A}$. For pedagogical reviews, see~\cite{rogachevskii2017laminar}.

As a consequence, the presence of a non vanishing $\textbf{E.B}$ term alters the otherwise expected chiral symmetry, giving a contribution to the magnetic energy. The presence of stellar electromagnetic fields within the star will induce a breaking of the spacial symmetry signaling a definite coupling direction and thus affecting the DM stellar distribution.

It is worth noting here that this coupling has been previously studied yielding observable phenomena, mostly in the NS atmosphere and beyond where the spatial field lines can be reasonably modelled. Such effects produce  axion-to-photon conversion \cite{millar2021axion} or axion-induced radiation in the presence of intense magnetic fields \cite{Xu_2022}.

In this work, we explore the spatial distribution of axion-like dark matter within a magnetized NS described in the General Relativistic (GR) framework under the presence of a strong, non-uniform poloidal magnetic field and a much weaker electric field $E\ll B$. For this, the stellar structure is modeled using a semi-analytical approach, incorporating the effects of the magnetic field whose strength has radial dependence, $B(r)$, inherited from its microscopic density dependence as baryonic density radial profile changes inside the NS from the high density inner core to the outer crust regions. Within the crust we will qualitatively describe the existence of flux tubes \cite{Shukla:2024nye} with enhanced intensity over an average field background value as possible local effective sites for axion-photon conversion.
We will assume temperature effects are such they do not appreciably alter the matter equation of state \cite{barbaPhysRevC.106.065806}. 
Specifically, it is already well known that B fields may quantize fermion dynamics \cite{landauPhysRevC.84.045803} and induce an anisotropic pressure from an assumed configuration, usually axisymmetric.  For magnetized NSs the assumed magnetic fields allowed by equipartition theorems predict as high as $\rm log(B[G])\lesssim 18$ and it turns out that, even then, the expected induced stellar deformations are tiny and thus are treated as a correction to the otherwise isotropic averaged total pressure. This result constitutes a validating proof to evaluate the axion stellar solutions constructed from original  stellar equilibrium configuration departing from a spherically symmetric solution.  

The manuscript is structured as follows in Section \ref{ii} we introduce the Lagrangians of interest with the axion and electromagnetic fields involved in the NS scenario to later compute the resulting stellar structure. Following \cite{Mallick}, we incorporate the effects of a density-dependent magnetic field on the equilibrium configuration. We will restrict axions coupling to electromagnetic fields leaving matter fields for a future contribution. In this sense some works \cite{lecce2025probingaxionlikeparticlesmultimessenger} have stated that while the production of ALPs in the hot and dense stellar nucleon scenarios depends on axion-nucleon coupling, the observational prospects in electromagnetic channels are governed by the axion conversion via coupling to fields in magnetized environments. 
We evaluate both the scalar ALPs (S-ALPs) and pseudoscalar ALPs (PS-ALPs) nature allowed, in principle, by the Standard Model. We obtain the dynamical equations solving for static configurations.  In Section \ref{iii} we detail the explicit formulation of the static axion field equations in two dimensions for the obtained stellar structure. In Section \ref{iv}, we discuss the axion field equation along with boundary conditions and find, first, the solution with no axion source, showing the explicit dependence of this solution with the mass of the field. In Section \ref{v} and \ref{vi}, we present the main results of our study for ALPs of scalar and pseudoscalar nature, highlighting the impact of their coupling  to electromagnetic fields on their spatial stellar distribution. In Section \ref{conversion} we explore on a tube-like distribution of fields in the crust and discuss the indirect observational prospects to finally conclude in Section \ref{conclusion}.

\section{Stellar Structure and axion field equations}\label{ii}

We first present the NS scenario and describe the stellar structure equations in presence of a  magnetic field  ${\bf B}$ and a much weaker electric field ${\bf E}\ll {\bf B}$ so that only the former will effectively play a role in global symmetry.  
Following the approach in \cite{Mallick}, we adopt a model that describes structural deformations caused by small deviations from spherical symmetry due to a dipolar magnetic field. In this approximation, both the metric tensor and the thermodynamic pressure are expanded in spherical harmonics $Y_{\ell m}(\theta, \varphi)$ up to quadrupole order.

The spherical harmonics are related to the associated Legendre functions $P_\ell^{m}(\cos \theta)$ and, for $m \geq 0$, they take the form

\begin{equation}
Y_{\ell m}(\theta, \varphi) = (-1)^m N_{\ell m} P_\ell^m(\cos \theta) e^{i m \varphi},
\end{equation}

with normalization factor $N_{\ell m} = \sqrt{\frac{2 \ell+1}{4 \pi} \frac{(\ell-m)!}{(\ell+m)!}}$. Monopole terms determine the radial dependence, while quadrupole terms introduce corrections to the stellar geometry, modifying both radial and angular dependencies.

From a microscopic perspective, inside the NS matter and fields can be described by the equation of state (EoS). Thus the energy density, $\varepsilon$, and the pressure, $P$, inside the star are  obtained, in a first approximation from the ordinary matter ($m$) i.e. typically composed of baryons, leptons, and other matter fields. Relying on more refined treatments matter magnetization is indeed possible, see \cite{perezPhysRevC.80.045804}, but for simplicity we are neglecting this contribution here.  

We first start by describing our NS benchmark model as an axion-free model such that the energy density reads

\begin{equation}
\varepsilon =\varepsilon_m+\frac{B^2}{8 \pi}, 
\end{equation}

and the pressure can be expressed in a Legendre polynomial expansion as 
\begin{equation}
   P=P_m+\left[p_0P_0(\cos \theta)+p_2 P_2(\cos \theta)\right]. 
\end{equation}
Explicitly, $p_0\equiv\frac{B^2}{24 \pi}$ is the monopole coefficient contribution and $p_2\equiv-\frac{ B^2}{6 \pi}$ the quadrupole coefficient contribution to the magnetic pressure. $P_0(x)=1$ and $P_2(x)=\frac{1}{2}\left(3 x^2-1\right)$ are the second-order Legendre polynomials associated to spherical harmonics  $m\equiv0$ subset fulfilling orthogonality relations. Thus for the  perpendicular (${\perp}$) and parallel (${\|}$) directions related to that of $\bf B$, pressure differences arise under the form

\begin{equation}
\begin{aligned}
P_{\perp} & =P_m+\frac{B^2}{8 \pi}, \\
P_{\|} & =P_m-\frac{B^2}{8 \pi}.
\end{aligned}
\end{equation}

Note that for the magnetic energy density the Centimeter–Gram–Second unit conversion used is $
\varepsilon\left[\mathrm{erg} / \mathrm{cm}^3\right]=\frac{1}{8 \pi}(B \text { [Gauss] })^2$. By using this expansion, the line element in stellar $(r,\theta, \phi)$ coordinates can be written as
\begin{equation}\label{metric_per}
d s^2=-e^{\nu(r)}(1+2h(r,\theta)) c^2d t^2+e^{\lambda(r)}(1+\frac{G m(r,\theta) e^{\lambda(r)}}{rc^2})d r^2+r^2(1+2k(r,\theta))\left(d \theta^2+\sin ^2 \theta d \varphi^2\right),
\end{equation}
 where in the same multipolar expansion up to quadrupole order we have
\begin{equation}\label{hmk}
\begin{aligned}
h(r,\theta) & =h_0(r)+h_2(r)P_2(\cos(\theta))+...,\\
m(r,\theta)&=m_0(r)+m_2(r)P_2(\cos(\theta))+...,\\
k(r,\theta)&=k_2(r)P_2(\cos(\theta))+...,
\end{aligned}
\end{equation}

with higher-order terms ($\ell \geq 4$) omitted for simplicity and $h_0(r), h_2(r), m_0(r), m_2(r), k_2(r)$ are $r$-dependent coefficients in the expansion.  $G$ is the gravitational constant and $c$ is the speed of light. The metric functions $\nu(r)$ and $\lambda(r)$ can be determined through 
\begin{equation}
\begin{aligned}
\frac{d \nu}{d r} & =-\frac{2}{\varepsilon_m+P_m} \frac{d P_m}{d r}, \\
e^\lambda & =\left(1-\frac{2 G m_{\rm sph}(r)}{rc^2}\right)^{-1},
\end{aligned}
\end{equation}  
being $m_{\rm sph}(r)$ the mass enclosed within the radius $r$ when $B=0$ and spherical symmetry holds. Note that if we fix for all $\ell$, $h_\ell=m_\ell=k_\ell=0$ this is equivalent to null B-field and we also recover the fully spherically symmetric Schwarzschild  metric.
From the Einstein equations and  the conservation laws as derived from the associated metric tensor, we can further obtain the energy density and pressure components\duvcom{,} so that after some mathematical transformations\duvcom{,} we obtain 
\begin{align}
    &\frac{d m_0}{d r}=4 \pi r^2 \frac{p_0}{c^2}, \label{mallick:m0}\\
    &\frac{d h_0}{d r}=4 \pi r e^\lambda \frac{G p_0}{c^4}+\frac{G}{rc^2} \frac{d \nu}{d r} e^\lambda m_0+\frac{G}{c^2 r^2} e^\lambda m_0,  \label{mallick:h0}\\
    &h_2+\frac{e^\lambda}{r} \frac{G m_2}{c^2}=0,\label{mallick:m2}\\
    &\frac{dk_2}{dr}=\frac{2 p_2 \frac{d\nu}{dr}+\frac{dp_2}{dr}}{\varepsilon + P},\label{mallick:conservation_1}\\
&\frac{dh_2}{dr}=\frac{-p_2 \frac{d\nu}{dr}-\frac{dp_2}{dr}}{\varepsilon + P}\label{mallick:conservation_2}.
\end{align}
In our setting this cold system is fully determined by specifying a matter EoS  $\varepsilon_m(P_m)$ and the magnetic field intensity $B$. Consequently, we begin by assuming that the internal magnetic field depends on baryon number density, $n$, following \cite{Mallick} and \cite{anzuiniPhysRevD.109.083030} in the form
\begin{align}\label{magneticprofile}
    B(n)=B_s+B_0 (1-e^{-\beta(\frac{n}{n_0})^{\delta}})
\end{align}
with $B_s$ and $B_0$ the value of the magnetic field at the surface and in the center of the star, respectively. Parameters $\beta=0.01$ and $\delta=2.0$ are fixed. 
$n_0$ is the baryonic number density at saturation $n_0\sim 0.15$ $\rm fm^{-3}$. The parameters $\delta$ and $\beta$ control how fast the average central magnetic field value $B_0$ decreases to that at the surface  $r=R$ as $B_s$. Observationally for  magnetars $B_s \sim 10^{14}-10^{15} ,\,\mathrm{G}$. The central magnetic field strength remains at present as quite uncertain quantity, although it is a reasonable assumption to consider it may be a factor $\sim10^2-10^3$ larger than in the surface. For the latter some cyclotron line have yielded experimentally deduced values around $10^{15}$ \cite{Ibrahim_2002}. In the stellar volume and from equipartition between the magnetic energy and the kinetic energy on the smallest scales, thus yielding fields of the order $\rm logB[G]>16$  local values \cite{Chabanov_2023} and based on more general theoretical grounds $B_0\sim 10^{18}-10^{19} \,\mathrm{G}$ are allowed  \cite{perezPhysRevC.84.045803}. 

In order to describe the interior baryonic EoS we will adopt a generic polytropic form, since it captures the features of the majority of the more sophisticated models when piece-wise procedures are used \cite{sulePhysRevC.104.015801}. Nevertheless, we note that the astrophysical constraints on mass and radius are satisfied in the cases studied in this work, while a more detailed analysis is left for future work.

On general grounds, the matter energy density $\varepsilon_m$ is calculated by using the first law of thermodynamics in the zero-temperature limit  in a  differential form $ d\left(\frac{\varepsilon_m}{\rho}\right)=-P_m d\left(\frac{1}{\rho}\right) $.

Thus the polytrope for matter description takes the form $P_m=K \rho^{\Gamma},$ with rest-mass density $\rho$, $K$ being the polytropic constant, and $\Gamma$ being the adiabatic index. This simplified EoS, while approximate, is commonly used in practice. The system closure is achieved with
\begin{align}
    \varepsilon_m(P)=\left(\frac{P_m}{K}\right)^\frac{1}{\Gamma}c^2+\frac{P_m}{\Gamma-1},
\end{align}

where the first term represents the rest mass contribution and the second one the internal energy.

\begin{figure}[H]
    \begin{subfigure}[H]{0.48\textwidth}  
        \includegraphics[
            width=\linewidth,  
        ]{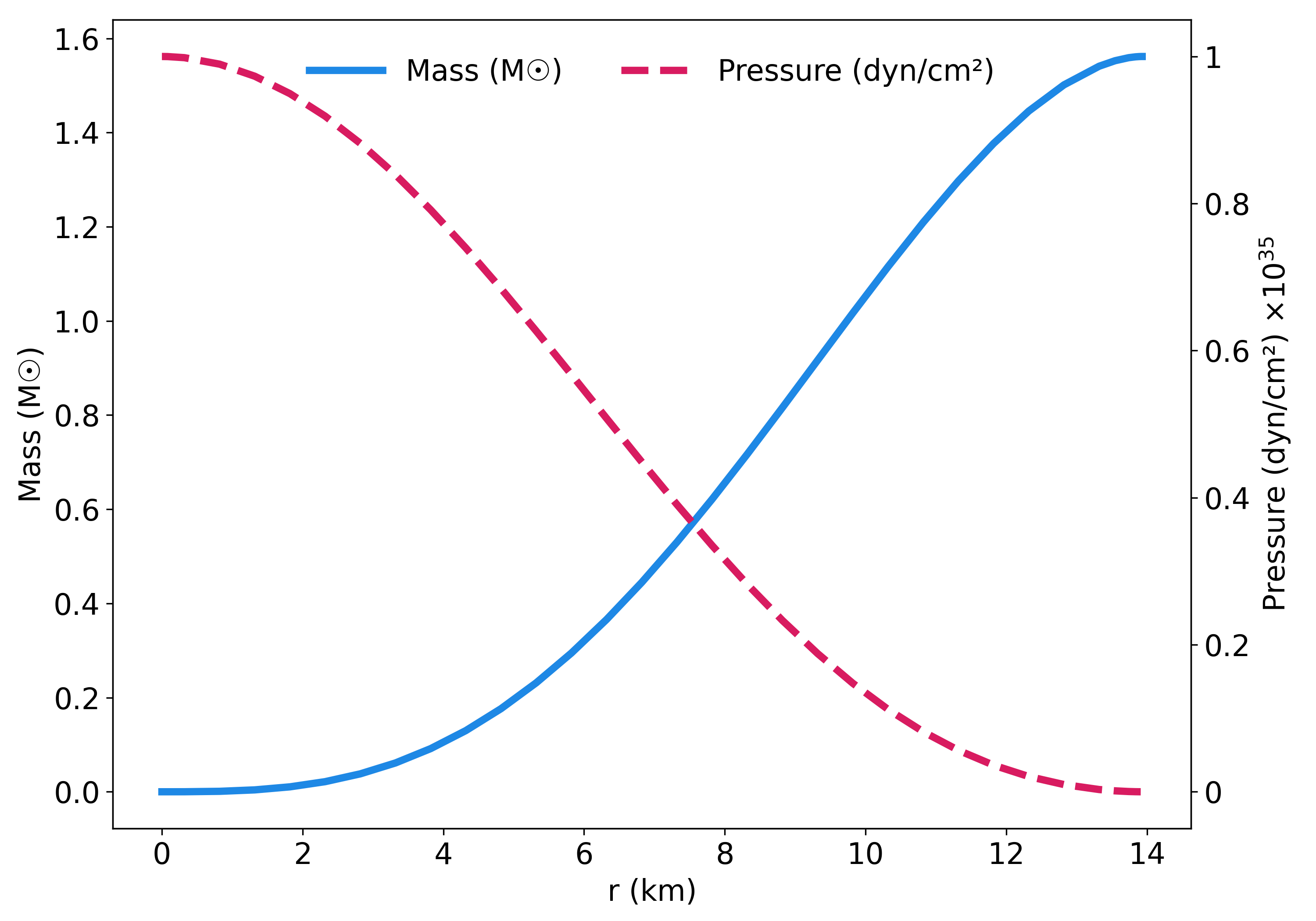}
    \end{subfigure}
    \begin{subfigure}[H]{0.48\textwidth}
        \includegraphics[
            width=\linewidth,  
        ]{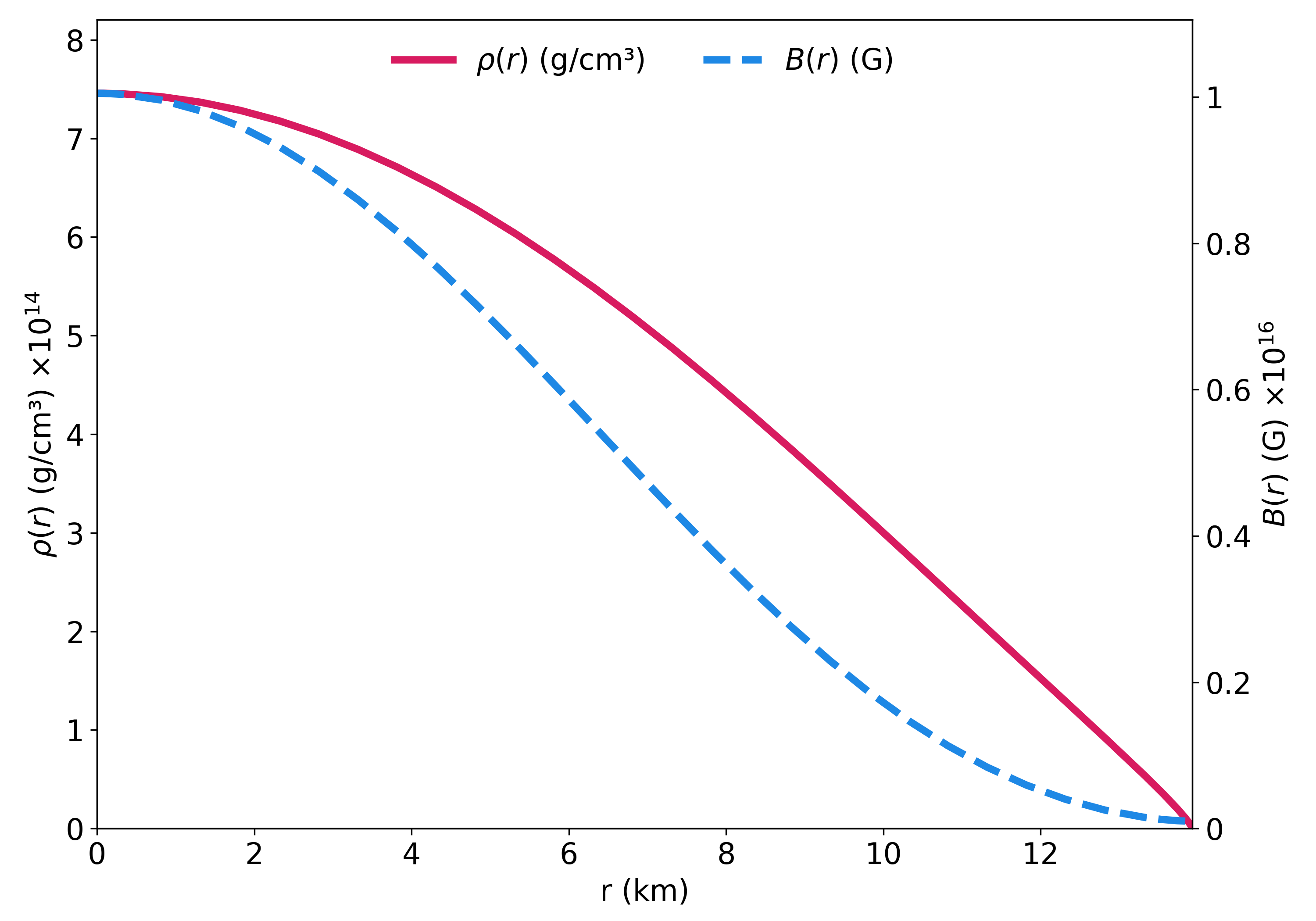}
    \end{subfigure}
      \caption{(left) Stellar mass and pressure radial profiles for a NS with $M= 1.56 M_{\odot}$ and $R=13.93 $ km (right) mass density and magnetic field strength radial profiles. We have used a polytropic EoS with $\Gamma=2.207$,$K=1.5\times 10^2$ $\mathrm{dyn} \cdot \mathrm{cm}^{3 \Gamma-2} \cdot \mathrm{~g}^{-\Gamma}$. B field is included under the prescribed form in Eq.\eqref{magneticprofile}, see text for details.}
    \label{fig1}
\end{figure}

Now solving Eqs.\eqref{mallick:m0}, \eqref{mallick:h0}, \eqref{mallick:m2}, \eqref{mallick:conservation_1} and \eqref{mallick:conservation_2} one can calculate the $h_0(r), h_2(r), m_0(r)$, $m_2(r)$, $k_2(r)$ in Eq. \eqref{hmk} as shown in \cite{Mallick}.  In Fig.\eqref{fig1} we show (left) the stellar mass and pressure radial profiles for a NS with $M= 1.56 M_{\odot}$ and $R=13.93126 $ km (right) mass density and magnetic field strength radial profiles. We have used a polytrope EoS with $\Gamma=2.207$, $K=1.5\times 10^2$ $\mathrm{dyn} \cdot \mathrm{cm}^{3 \Gamma-2} \cdot \mathrm{~g}^{-\Gamma}$. This particular set of values $(M,R)$ lies within the inner $68\%$ confidence region  obtained for  PSR J0030+0451 mass and radius estimates from NICER Data \cite{Miller_2019}. B field is included as prescribed in Eq. \eqref{magneticprofile} using $B_0=10^{18}$ G, $B_s=10^{14}$ G and $\beta=0.01, \delta=2.0$. 

\section{Axion fields inside the Neutron Star}\label{iii}

We assume that ultralight axion-like particles (ALPs) populate the interior of the neutron star, contributing negligibly to the energy–momentum tensor compared to ordinary matter. More generally, we allow for ALPs by using light scalar or pseudoscalar fields differing from the QCD axion originally motivated to solve the strong-CP problem. At this point a remark is due regarding the scenario we consider. In this work the ALP distribution inside the star is subject, mainly, to gravitational and electromagnetic interactions, with subleading coupling to ordinary matter fields. In this sense we consider an internal stellar finite density of ALPs to be essentially photophilic and mostly stabilized due to highly magnetized and compact stars. 

Thus we will consider the NS structure and the surrounding spacetime can be treated as a fixed background, determined solely by the EoS of baryonic  matter and the metric of GR curved space-time including the stellar magnetic field, {\bf B} under the form in Eq. \eqref{metric_per}. Note we consider that the associated electric field {\bf E} will play no role in structure as, at best, will be orders of magnitude smaller, ${E}\ll B$, as a result of tiny chiral imbalance and  weak beta equilibrium conditions \cite{Gonalves2022}.

In order to derivate the static solution to the axion field as a function of stellar radial and azimutal coordinates, $a(r,\theta)$ we now present the associated Lagrangian density for the scalar (S) and pseudoscalar (PS) nature of the axions we consider in this work.  The axion-photon interacting system is described by the generic Lagrangian

\begin{align}\label{lagg2}
\mathcal{L} &= \frac{1}{2} \partial_\mu a \partial^\mu a - \frac{1}{2}m^2a^2 - \frac{1}{4\mu_0}F_{\mu\nu}F^{\mu\nu} + \mathcal{L}_{\text{int}}, \\
m &\equiv \frac{c}{\hbar}m_a, \quad F_{\mu\nu} \equiv \partial_\mu A_\nu - \partial_\nu A_\mu \nonumber,
\end{align}

where $a$ is the axion field and $m_a$ its mass. Here the potential energy term is $\varepsilon_{\rm axion}=\frac{1}{2}m_a^2 a^2$. $F_{\mu\nu}$ is the electromagnetic  field tensor strength. The electric and magnetic field components are obtained as $E^i= F^{i0} $ and $ B^i = -\frac{1}{2}\epsilon^{ijk}F_{jk} = \tilde{F}^{i0}$
with dual tensor $\tilde{F}^{\mu\nu} \equiv \frac{1}{2}\epsilon^{\mu\nu\rho\sigma}F_{\rho\sigma}$ using the 4th rank Levi-Civita tensor. The fundamental Lorentz and gauge invariants \cite{Escobar_2014} are convenient quantities to use in this context, namely $F_0 \equiv -\frac{1}{4}F_{\mu\nu}F^{\mu\nu} = \frac{1}{2}(\mathbf{E}^2 - \mathbf{B}^2)$ and $G_0 \equiv -\frac{1}{4}F_{\mu\nu}\tilde{F}^{\mu\nu} = \mathbf{E} \cdot \mathbf{B}$. From this point on we use Lorentz-Heaviside units with $\mu_0=1$.
Inspired by \cite{Paix_o_2022}, we consider, generically, the S or PS axion-photon interaction Lagrangian can be written under the form

\begin{align}
&\mathcal{L}_{\text{int}} = -{g_{a,\gamma}}Q(F_0,G_0), \\
&Q(F_0,G_0) \equiv 
\begin{cases}
G_0 & \text{(pseudoscalar coupling)} \\
F_0 & \text{(scalar coupling)}
\end{cases}
\end{align}
where we will be restricting to only linear contributions in $Q$ i.e. $F_0,G_0$ for the sake of simplicity. Non linear couplings are indeed possible and have been somewhat explored \cite{mscmapa2014}. 
Our focus here is on the influence of these strong magnetic fields and the associated invariant interaction Lagrangians on the spatial distribution of axions.For this we consider only the axion–photon coupling as a first approximation, having in mind, however, matter-ALP interaction should not be neglected in dense scenarios \cite{anzuiniPhysRevD.109.083030,kumar2024cp, Buschmann_2022,Lella_2024,cabo2021dynamical} 

Returning to the electromagnetic case, let us mention here that, in the pseudoscalar case the allowed coupling of the ALP field, $a$, to the electromagnetic field arises via the Lagrangian $\mathcal{L}_{a\gamma} = g_{a\gamma} \, a \, \mathbf{E} \cdot \mathbf{B},$ originating from the Lorentz-invariant electromagnetic tensor contraction 
$\mathcal{L}_{\text{int}} = \frac{1}{4} g_{a\gamma} \, a \, F_{\mu\nu} \tilde{F}^{\mu\nu}.
$ This coupling is CP-even provided that  $a$ is CP-odd, and governs key physical processes such as ALP-photon conversion in external magnetic fields as we will explain later in this manuscript. On the other hand, if the ALP is considered to be a scalar (CP-even), then the CP-conserving interaction must instead involve the term $ \mathcal{L}_{a\gamma} = g_{a\gamma} \, a \, (\mathbf{E}^2 - \mathbf{B}^2),
$ which arises from the invariant combination
$\mathcal{L}_{\text{int}} = \frac{1}{4} g_{a\gamma} \, a \, F_{\mu\nu} F^{\mu\nu}$ \cite{ballou2015new}.
This distinction is crucial, as each type of coupling leads to qualitatively different dynamics and subsequent properties.

From the Lagrangian in Eq.\eqref{lagg2} we obtain the axion follows the Klein-Gordon equation with the electromagnetic field source
\begin{align}
\Box a + m_a^2a &= {g_{a \gamma}}Q,
\label{eq:dalambertian}
\end{align}
where the d'alembertian operator can be written as $\Box a \equiv \frac{1}{\sqrt{-g}}\partial_\mu\left(\sqrt{-g}g^{\mu\nu}\partial_\nu a\right)$. So 
 the axion field configuration in this magnetized spacetime is governed by a modified Klein-Gordon equation expressed in polar coordinates $(r,\theta)$, where the magnetic field exerts influence not only through the explicit interaction term but through the metric coefficients $f(r,\theta)$ and $k(r,\theta)$. A detailed explanation is given in Appendix \eqref{appendix1}. The governing equation Eq.\eqref{eq:dalambertian} adopts the form 

\begin{align}\label{axion_general}
    \frac{1}{f}\partial_r^2 a+\frac{1}{r^2k}\partial_\theta^2 a+\tilde{\xi_1}\partial_r a+ \tilde{\xi_2}\partial_\theta a +m^2a =g_{\rm eff} Q(F_0,G_0),
\end{align}
where we have defined $g_{\rm eff}\equiv \frac{g_{a \gamma}}{ a_0}$ and $a_0$ is the stellar central axion field value to be later determined by setting an axion stellar density. Here, $f(r,\theta)$ and $k(r,\theta)$ encode spacetime curvature effects from the stellar magnetic field.  $\tilde{\xi}_1(r,\theta),\tilde{\xi}_2(r,\theta)$ contain first-derivative terms being 
\begin{align}
&\tilde{\xi}_1(r,\theta)\equiv\frac{\xi_1(r,\theta)}{f(r,\theta)}-\frac{f'(r,\theta)}{f(r,\theta)^2}, ~ ~ ~~~~\xi_1(r,\theta)\equiv\frac{2}{r}+\frac{h'}{2h}+\frac{f'}{2f}+ \frac{k'}{k},\\
&\tilde{\xi}_2(r,\theta)\equiv\frac{\xi_2(r,\theta)}{r^2 k(r,\theta)}-\frac{\oth{k}(r,\theta)}{r^2k(r,\theta)}, ~ ~ \xi_2(r,\theta)\equiv\cot{\theta}+\frac{\oth{h}}{2h}+\frac{\oth{f}}{2f}+\frac{\oth{k}}{k},
\end{align}
with the notation $a'\equiv\partial_r a$ and $\oth{a}\equiv\partial_\theta a$.

\subsection{Free axion fields $g_{\rm eff}=0$} 
\label{iv}

We start our analysis with the most simplified case where the axion field does not interact with the stellar electromagnetic field directly, i.e. $g_{\rm eff}=0$. {In this case the solution $a(r,\theta)$ only depends on the magnetic fields through the background metric}. In this way we have to solve 

\begin{align}\label{axion2d}
    \frac{1}{f(r,\theta)}\partial_r^2 a(r,\theta)+\frac{1}{r^2k(r,\theta)}\partial_\theta^2 a(r,\theta)+\tilde{\xi_1}(r,\theta)\partial_r a(r,\theta)+ \tilde{\xi_2}(r,\theta)\partial_\theta a(r,\theta) +m^2a(r,\theta) =0.
\end{align}
 This is a  boundary value problem that we  specify through the following conditions
 \begin{align}
     &a(0,\theta)=a_0,\\
     &a(r_{\rm surface},\theta)=a_{\rm sph}(r_{\rm surface}),\\
     &a(r,0)=a(r,\pi)=a_{\rm sph}(r).
 \end{align}

We consider the central axion field value, $a_0$ as a parameter of our problem where stellar axial symmetry induces prolate (oblate) surface radii as $r_{\rm surface}$ that we fix to the equivalent spherical value  $a_{\rm sph}$ obeying 
 \begin{align}
    &r_{surface} = \frac{R_e R_p}{\sqrt{R_p^2 \sin^2\theta + R_e^2 \cos^2\theta}},\\
    &\partial_r^2a_{sph}(r)+\frac{1}{2}(\frac{4}{r}+\nu'(r)-\lambda'(r))\partial_ra_{sph}(r)+m^2e^{\lambda(r)}a_{sph}(r)=0\label{asph}.
\end{align}
We consider $R_e$ as the equatorial radius and $R_p$ the polar radius, following \cite{Mallick}.
This two radii can be calculated as a departure from spherical value  $R=R_{\rm sph}+\delta R$ with  $\delta R \equiv \psi_0(R) +(\psi_2(R)+Rk_2(R))P_2(\cos{\theta})$ \cite{rizaldy2018magnetized}
being
\begin{align}
    \psi_0(r)\equiv-\frac{r(r-2Gm(r))}{G(4\pi r^3 P_m + m(r))}h_0(r),\\
    \psi_2(r)\equiv-\frac{r(r-2Gm(r))}{G(4\pi r^3 P_m + m(r))}h_2(r),
\end{align}
where $h_0(r),h_2(r)$ are obtained from Eq. \eqref{mallick:conservation_2}. Finally
\begin{align}
    &R_e\equiv R_{sph}+\psi_0(R)-\frac{1}{2}(\psi_2(R)+Rk_2(R)),\\
    &R_p\equiv R_{sph}+\psi_0(R)+(\psi_2(R)+Rk_2(R)).
\end{align}
This formulation allows the systematic study of magnetic field-induced deformations on axion profiles through the metric coefficients, while maintaining connection to known spherical solutions via boundary matching \cite{Mallick}.\\

Before proceeding with the solution to Eq.\eqref{axion2d}, it is important to discuss the boundary condition outlined in Eq.\eqref{asph}. This equation formally resembles that of a spatially damped oscillator, where the effective damping term is influenced by the metric components, which, in turn, are determined by the background matter EoS. Thus, even if we assume that the interactions between the axionic field and baryonic matter are sufficiently weak to be neglected, the behavior of the system is still indirectly affected by the EoS. 

On the other hand, in a similar fashion to damping behaviour,  different regimes in the $a(r,\theta)$ solution, are determined by the value of the quantity

\begin{equation}
    \Delta(r)=\frac{1}{4}(\frac{4}{r}+\nu'(r)-\lambda'(r))^2-4m^2e^{\lambda(r)},
\end{equation}
thus, for values of the radius sufficiently close to the center of the star, the term in the first parenthesis $\sim\frac{4}{r}$ dominates,  driving the system to $\Delta > 0$, and the regime becomes overdamped, causing the value of the axion field to decrease without oscillation. Depending on the axion mass, this region will be closer or farther from the center. The lower the ALP mass, the more external the NS region is where the axionic field value decays. 
\begin{figure}[H]
       \begin{center}
    \begin{subfigure}[b]{0.47\textwidth} 
        \centering
        \includegraphics[width=8.5cm, height=6.7cm]{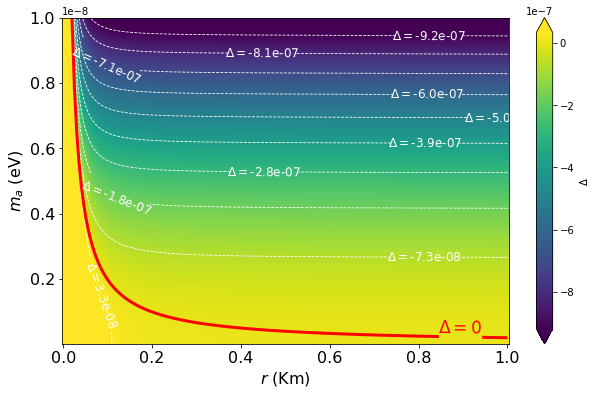}
        \label{fig:delta}
    \end{subfigure}
    \hfill 
       \begin{subfigure}[b]{0.47\textwidth}
        \includegraphics[width=8.5cm, height=6.5cm]{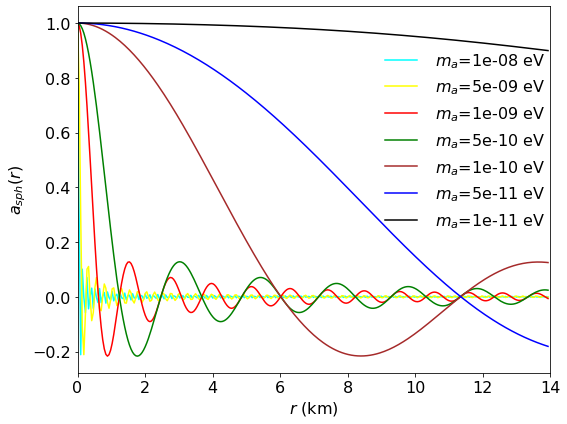}
        \label{fig:boundary_condition}
    \end{subfigure}
    \end{center}
   \caption{(right) Axion field spherically equivalent $a_{\rm sph}(r)$ solutions for different  axion masses in the range $m_a\in [ 10^{-11},10^{-8}]$ eV. (left) $\Delta$ values in the inner stellar kilometer. The red line marks the isocontour where $\Delta$ vanishes. Below the red line the axion field is overdamped while above the axion field is  underdamped. 
    }
    \label{fig:axion1D}
\end{figure}

The overdamped regime ($\Delta > 0$) near the stellar core indicates strong metric-induced damping, where spacetime curvature due to baryonic density minimizes axion effects. This mimics a screening mechanism. For ``heavy'' axions, the mass term $4m^2e^{\lambda(r)}$ dominates, shrinking the overdamped region and allowing underdamped oscillations ($\Delta < 0$) to develop closer to the stellar core. 
In Fig. \eqref{fig:axion1D} we plot (right panel) spherically equivalent $a_{\rm sph}(r)$ solutions  to Eq. \eqref{asph} for the same stellar configuration varying axion masses $m_a\in [ 10^{-11},10^{-8}]$ eV. We see that as the axion mass increases, the characteristic damping scale of the solutions decreases, establishing an upper mass limit $m_a\sim 10^{-10}$ eV for halo formation around the star. Axions with masses exceeding this value cannot support halo solutions in magnetized NS as they will mostly be contained inside the stellar volume. On the left panel in Fig. \eqref{fig:axion1D} we show the color scale value of $\Delta$ for the same range of axion mass as in the right panel constrained to the core innermost kilometer. The red line marks the isocontour where the value $\Delta=0$. $\Delta$ values below the red line are positive and, as a consequence, the axion field is overdamped. Instead $\Delta$ values above the red line are negative and will result in solutions where the field will be underdamped rapidly oscillating to null values inside the stellar volume. 

As a result, we can define $r_c$ as the critical radius where the damping regime changes and becomes determined, given an axion mass value, by the red line boundary. In the mass range under study and for heavy axions $r_c\ll R_e$ resulting in a small axion core while for light axions overdamping causes their distribution extending over regions ($r_c \sim R_e$), where the field smoothly decays monotonically throughout most of the stellar volume.

In order to solve Eq. \eqref{axion2d} we have taken the following conditions for axion-photon null interaction case $g_{\rm eff}\equiv0$. 
We select a central pressure ($p_c=1\times10^{35}$ $\rm dyn \,cm^{-2}$), and for the EoS we fix  $\Gamma=2.207$,  $K=1.5\times 10^2$ $\mathrm{dyn} \cdot \mathrm{cm}^{3 \Gamma-2} \cdot \mathrm{~g}^{-\Gamma}$. We take $m_a=10^{-10}$ eV.  The magnetic field  profile is taken by fixing in Eq. \eqref{magneticprofile} along with the parameters $\beta=0.01$ and $\delta=2.0$. In addition, the central value of the magnetic field is $B_0=10^{18}\,\rm G$ while  in the surface $B_s=10^{14}\,\rm G$. Regarding the spherically equivalent star we obtain $M_{\rm sph}=1.56 M_{\odot}$ and $R_{\rm sph}=13.93$ km in the non-perturbed case.

\begin{figure}[H]
    \centering
    \begin{subfigure}[b]{0.49\textwidth} 
        \centering
        \includegraphics[width=\textwidth]{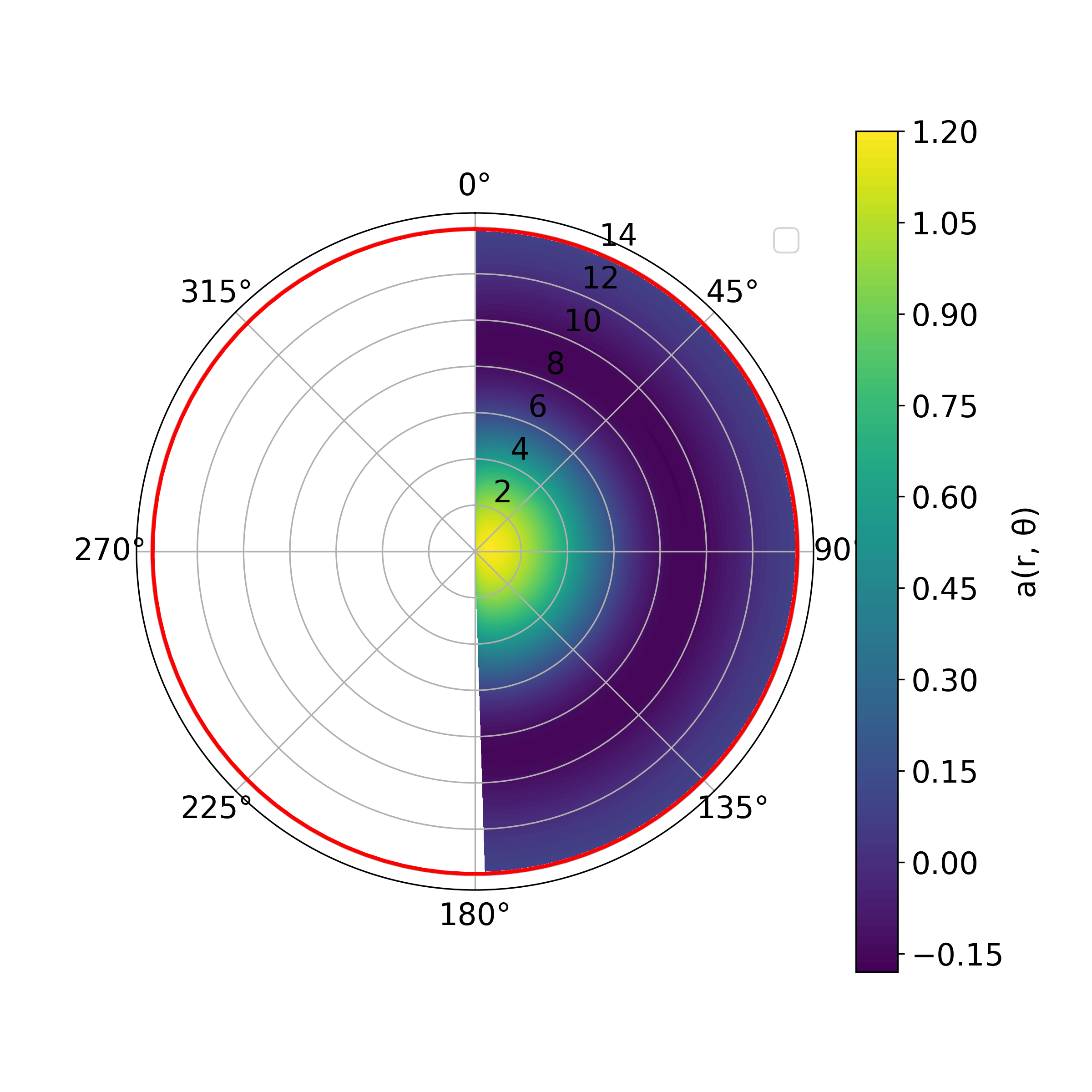} 
        \label{fig:subfig1}
    \end{subfigure}
    \begin{subfigure}[b]{0.49\textwidth}
        \centering
        \raisebox{1.0cm}{\includegraphics[width=\textwidth]{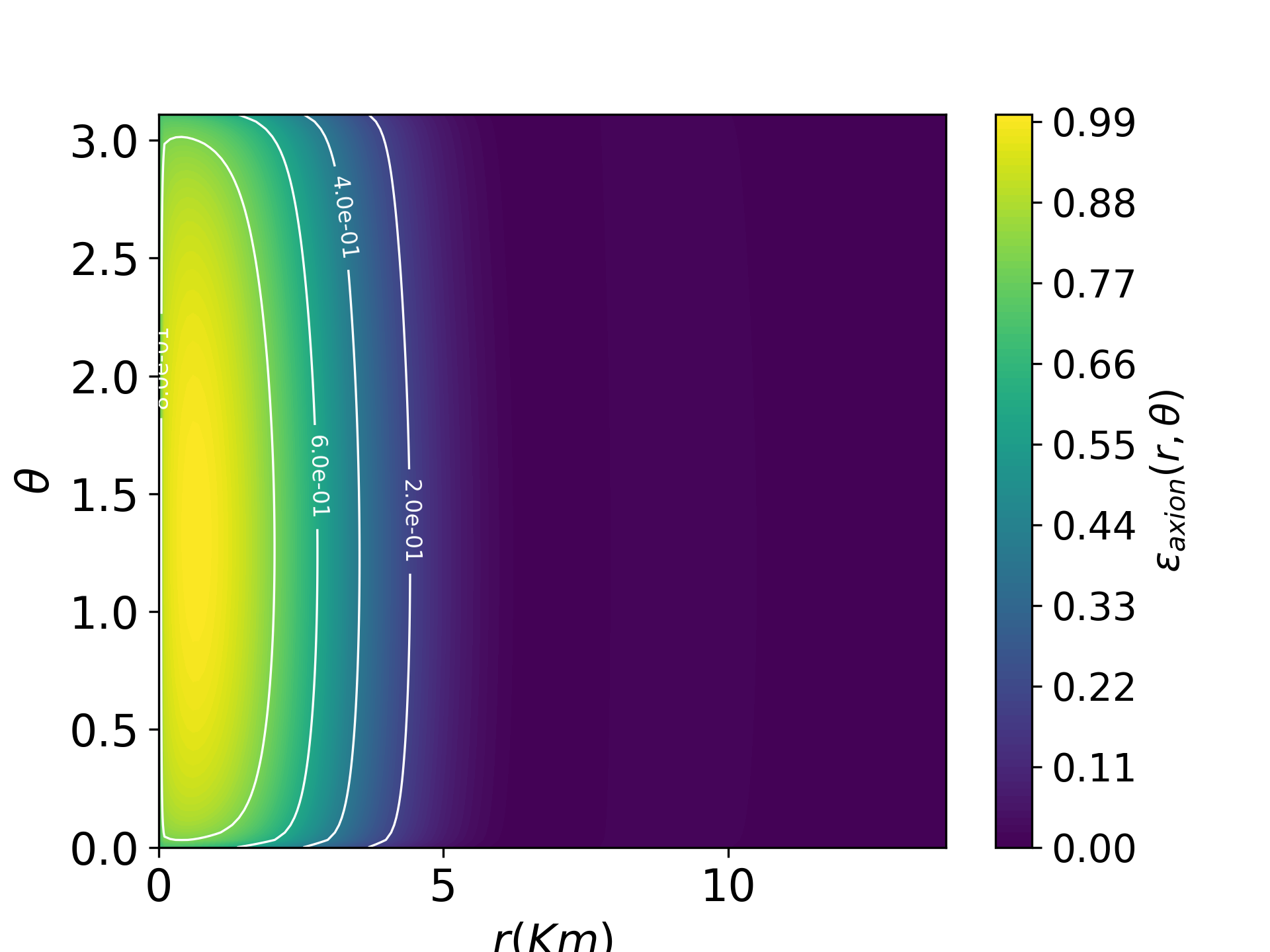}} 
        \label{fig:subfig2}
    \end{subfigure}
    \caption{(left) Internal axion field distribution in curved and magnetized stellar space-time with no coupling to electromagnetic fields, i.e. $g_{a,\gamma}=0$ and a mass $ m_a = 10^{-10} \, \text{eV}$ presented in an axial cross-section view along the polar diameter. (right) Potential energy $\varepsilon_{\rm axion}$.  The NS has a mass $ M_{\text{sph}} = 1.56 \, M_\odot $ and a radius $ R = 13.93 \, \text{km} $. The total axion mass was taken to be $ 0.01\% $ of the total NS baryonic mass.
  }
    \label{fig:figura_principal}
\end{figure}

The solution to the axion field equation Eq. \eqref{axion2d} yields the axion distribution shown in Fig.  \eqref{fig:figura_principal}. The left panel shows the amplitude $a(r,\theta)$ in a meridional plane while the right panel show its potential energy $\varepsilon_{\rm axion}=\frac{1}{2}m_a^2 a^2$. Note that in this case axions are symmetrically distributed around the center of the star, decaying rapidly with increasing stellar radius. If we now focus on the induced  magnetic field deformations, see appendix \eqref{appendix1} Eqs. \eqref{metric_h},\eqref{metric_f} and \eqref{metric_k} this solution describes a deformed star with an eccentricity of $\epsilon\equiv\sqrt{1-\left(\frac{R_p}{R_e}\right)^2}=5.21\times10^{-3}$ related with the ellipticity through $e\equiv \frac{1}{2}\epsilon^2=1.36\times10^{-5}$ and a total mass of $M=M_{\rm sph}+\delta M = 1.56+  4.52 \times 10^{-7} M_{\odot} $. If continuous gravitational waves are sourced by such an equatorial elliptiticty $\epsilon=\left|I_{x x}-I_{y y}\right| / I_{z z}$, with $I_{z z}$ being the moment of inertia of the star with respect to the principal axis aligned with the rotational axis, for this particular stellar configuration the gravitational wave strength can be obtained by\cite{Covas_2025} $h_0=10^{-26}\left[\frac{I_{z z}}{10^{38} \mathrm{~kg} \mathrm{~m}^2}\right]\left[\frac{e}{10^{-6}}\right]\left[\frac{f_0}{100 \mathrm{~Hz}}\right]^2\left[\frac{1 \mathrm{kpc}}{d}\right]$ yielding $h_0\sim10^{-25}$ within reach of 3rd generation detectors \cite{abac2025scienceeinsteintelescope}.

\subsection{Pseudoscalar axion field coupling  $Q(F_0,G_0)\equiv G_0\equiv g_{\rm eff}\textbf{E.B}$ }
\label{v}

In the ultradense interior of NSs ordinary matter  exhibits extraordinarily high electrical conductivity with values up to  $\sigma \sim 10^{23}\, \mathrm{s}^{-1}$ \cite{harutyunyan2024electrical}, which effectively suppresses the appearance of macroscopic electric fields by screening and neutralizing charge density perturbations on short timescales. However, when analyzing axion production in such environments, the source terms in the axion field equations, see Eq.\eqref{axion_general},  retain particular significance despite the mentioned screening. Crucially, these contributions depend not only on electromagnetic effects but also on the axion field density, parametrized by its central value amplitude $a_0$. \\
In what follows we will focus on the pseudoscalar axion or PS-ALP and solve the inhomogeneous Eq.\eqref{axion_general}, where the source term can be expressed as $Q(\textbf{E,B}) \equiv g_{\rm eff}\textbf{E.B}$ where $g_{\rm eff}$ is the effective coupling constant that encapsulates the intrinsic axion-photon coupling including a normalization due to the axion field value at the origin i.e $g_{\rm eff}\equiv \frac{g_{\gamma a}}{ a_0}$. 
We will  assume a stellar magnetic field in the cartesian $z$-direction whose magnitude depends on the density profile under the prescribed form in Eq.\eqref{magneticprofile} and a tiny but non-vanishing stellar electric field. 

Given the assumed NS scenario we now use the spherical unit-vectors  $\{\boldsymbol{\hat{e}_r}$ to express $\boldsymbol{\hat{e}_\theta},\boldsymbol{\hat{e}_\phi}\}$ spherical coordinate unit vectors as $\textbf{B}=B(\rho)\boldsymbol{\hat{k}}=B(\rho)\cos(\theta)\boldsymbol{\hat{e}_r}-B(\rho)\sin(\theta)\boldsymbol{\hat{e}_\theta}$ and, accordingly, the electric field as $\textbf{E}=\epsilon_r \boldsymbol{\hat{e}_r}+\epsilon_\theta \boldsymbol{\hat{e}_\theta}+\epsilon_\phi\boldsymbol{\hat{e}_\phi}$ with  $\epsilon_r,\epsilon_\theta, \epsilon_\phi \rightarrow0$ infinitesimal quantities. This leaves us with

\begin{align}
Q(\textbf{E,B})=g_{\rm eff}\textbf{E.B}&=g_{\rm eff}B(\rho)(\epsilon_r\cos(\theta)-\epsilon_\theta \sin(\theta)),
\end{align}
or in a reduced form
\begin{align}
    g_{\rm eff}\textbf{E.B}=    g_{\rm eff}\epsilon B(\rho),
\end{align}
 in the case $\epsilon_r= \epsilon \cos\theta$ and $\epsilon_\theta=-\epsilon\sin\theta$ as the electromagnetic fields are projected in the $z$-direction. 
 
 Note that, as already explained in Section {\eqref{Intro} , this is plausible if we assume the existence of  Chiral Magnetic Effects (CME) as in \cite{dvornikov2019permanent}. Thus there is a non-vanishing chiral number $N_5= \int d^3\mathbf{r} \, \bar{\psi} \gamma_5 \psi$, where $\psi$ represents massless spin-$1/2$ fermions. Chirality relaxation occurs via spin-flip processes and finite fermion masses, crucial in warm systems relaxing as the system cools down~\cite{Peskin:1995ev,sigl2016JCAP...01..025S}. The axial anomaly modifies current conservation laws as
 
 \begin{equation}
     \partial_\mu j_5^\mu = \frac{g^2}{32\pi^2} F_{\mu\nu}^\alpha \tilde{F}^{\alpha,\mu\nu} = \partial_\mu K^\mu ,
 \end{equation}
 where $K^\mu$ is the Chern-Simons current. This leads to the conserved quantity $\frac{d}{dt}\left(N_5 - N_{CS}\right) = 0$  where $N_{CS} \equiv \int d^3\mathbf{r} \, K^0$ is the Chern-Simons number.
For electromagnetic fields, $N_{CS}$ connects to the magnetic helicity,  $\mathcal{H}$, through $N_{CS} = \frac{e^2}{4\pi^2} \mathcal{H}$, where $\mathcal{H} \equiv \int d^3\mathbf{r} \, \mathbf{B} \cdot \mathbf{A}$,
yielding the helicity-chirality conservation law $\frac{d}{dt}\left(N_5 - \frac{e^2}{4\pi^2} \mathcal{H}\right) = 0.$
Therefore, chirality imbalance drives a current density 
 \begin{equation}
 \mathbf{J}_{\rm{CME}} = \frac{e^2}{4\pi^2} \mu_5 \mathbf{B}.
  \end{equation}
An associate chiral chemical potential arises as a result, $\mu_5 \equiv \frac{\mu_L - \mu_R}{2}$ depending on different lepton fractions with left (L) and right (R) chirality. 

We note at this point that while the CME prescribes a non-vanishing $\bf {E.B}$ source term can emerge in the system, there is a yet another, different contribution providing corrections adding to the previous.

Due to the CME, a non-vanishing electric field purely in the $z$-direction was taken into account. However, one must note that a component of it, contained in the meridional plane may arise as well, leading to a misalignment with $\textbf{B}$. This deviation can be induced by quantum effects, particularly what is known as the Chiral Vortical Effect (CVE) \cite{dvornikov2019permanent}. The CVE arises from the coupling between fluid vorticity and chiral imbalance, generating currents (not directly an electric field) proportional to the vorticity. These currents, in turn, can influence the electric field through Maxwell’s equations or resistive effects. This becomes particularly relevant when the magnetic field and fluid vorticity are misaligned, enhancing the emergence of a meridional electric field component.

In chiral Magneto-Hydrodynamics (chiral MHD), the electric field is modified indirectly through corrections to the current density. The CVE introduces a current term proportional to the vorticity $\boldsymbol{\omega}$, which alters the relationship between $\textbf{E}$ and $\textbf{B}$ in the generalized Ohm’s law. This modifies the effective electric field structure, potentially generating a meridional component and reshaping the system’s electrodynamics. These effects extend beyond ideal MHD, where the electric field is typically tied to the fluid velocity via $\boldsymbol{E} = -\boldsymbol{v} \times \boldsymbol{B}$ in the ideal limit. 

Therefore the generalized Ohm’s law in chiral MHD can be expressed as \cite{dvornikov2019permanent}

\begin{equation}\label{E_general}
    \mathbf{E}= \frac{c}{4\pi\sigma} \nabla \times \mathbf{B} - \frac{1}{c}\mathbf{v} \times \mathbf{B} + \frac{e^2}{2\pi^2\sigma \hbar^2 c} \mu_5\mathbf{B} + \frac{e\mu_e}{2\pi^2\sigma \hbar^2 c^{2}}\mu_5 \boldsymbol{\omega},
\end{equation}

where the last term, $\frac{e\mu_e}{2\pi^2\sigma \hbar c^2}\mu_5 \boldsymbol{\omega}$, directly relates the vorticity with the electric field. $e,\mu_e$ are the electron charge and chemical potential, respectively. In a NS in a rigid rotation approximation, the vorticity is predominantly aligned with the spin angular velocity $\mathbf{\Omega}\parallel \boldsymbol{\omega}$. However, when the magnetic axis is misaligned with the rotation axis, as observed in most pulsars and magnetars, the vectors $\mathbf{B}$ and $\boldsymbol{\omega}$ do not share the same orientation. This results in a nontrivial projection of $\mathbf{B}$ onto $\boldsymbol{\omega}$.
Consequently, the electric field $\mathbf{E}$ in the meridional plane does not necessarily align with $\mathbf{B}$. Instead, the presence of the chiral term $\sim \mu_5 \boldsymbol{B}$ introduces an additional component that modifies this alignment. 

In brief, assuming a NS where the spin and magnetic axis are not perfectly aligned, the projection of $\mathbf{B}$ onto  $\boldsymbol{\omega}$ alters the scalar product $\mathbf{E} \cdot \mathbf{B}$ in the meridional plane. This leads in the case under study to  $\epsilon_r \ne \epsilon_\theta$ so this become relevant even if its is small compared with $\frac{e^2}{2\pi^2\sigma} \mu_5\mathbf{B}$. In this case we can express the source term $Q(\textbf{E.B})$ as
\begin{align}
    g_{\rm eff}\boldsymbol{E.B}= g_{\rm eff} \frac{e^2}{2\pi^2\sigma \hbar^2 c} \mu_5B^2(r) + g_{\rm eff}\frac{e\mu_e}{\pi^2\sigma \hbar^2 c^2}\mu_5 \Omega B(r)\cos(\chi),
\end{align}
or equivalently
\begin{equation}
    g_{\rm eff}\boldsymbol{EB}= \alpha_1 B^2(r)  + \alpha_2 B(r)\cos(\chi),
\end{equation} 
where $\chi$ is the angle between $\mathbf{B}$ and  $\boldsymbol{\omega}$ directions and with 

\begin{equation}
     \alpha_1=g_{\rm eff}\frac{e^2\mu_5}{2\pi^2\sigma \hbar^2 c} ,
\end{equation}      
\begin{equation}
\alpha_2=g_{\rm eff}\frac{e\mu_e\mu_5 \Omega}{\pi^2\sigma \hbar^2 c^2},
\end{equation} 
and in consequence the axion Eq. (\eqref{axion_general} becomes now
\begin{align}\label{axion2d_withS}
    &\frac{1}{f(r,\theta)}\partial_r^2 a(r,\theta)+\frac{1}{r^2k(r,\theta)}\partial_\theta^2 a(r,\theta)+\tilde{\xi_1}(r,\theta)\partial_r a(r,\theta)+ \tilde{\xi_2}(r,\theta)\partial_\theta a(r,\theta) +m^2a(r,\theta) \nonumber \\ &=    \alpha_1 B^2(r) + \alpha_2B(r)\cos(\chi).
\end{align}

In order to size the relative contributions to the source term in what follows we will be considering a static frame so that $\Omega = 0$, besides average values for 
$\sigma \sim 10^{23}$ s$^{-1}$  as electrical bulk NS conductivity \cite{harutyunyan2024electrical, harutyunyan2016electrical} and chiral chemical potential $\mu_5 = 1.60218 \times 10^{-4}$ $\rm erg$, $\mu_e=1\times 10^{-4}$ erg in accordance to \cite{dvornikov2017relaxation}. Later on we will carefully explore the parameter space regarding prospects of observable signatures of axion condensates.
 As one can easily verify, for the non-rotating and millisecond pulsars cases with typical $\Omega \sim 100 \, \text{s}^{-1}$, it is consistently obtained  $\alpha_2\ll \alpha_1$. Therefore we conclude its contribution will be small compared to the term related to the CME and thus we discard it in what follows retaining only $\alpha_1$ dependence.
 
\begin{figure}[h]
    \begin{minipage}[b]{0.48\textwidth}
        \centering
        \includegraphics[width=\linewidth]{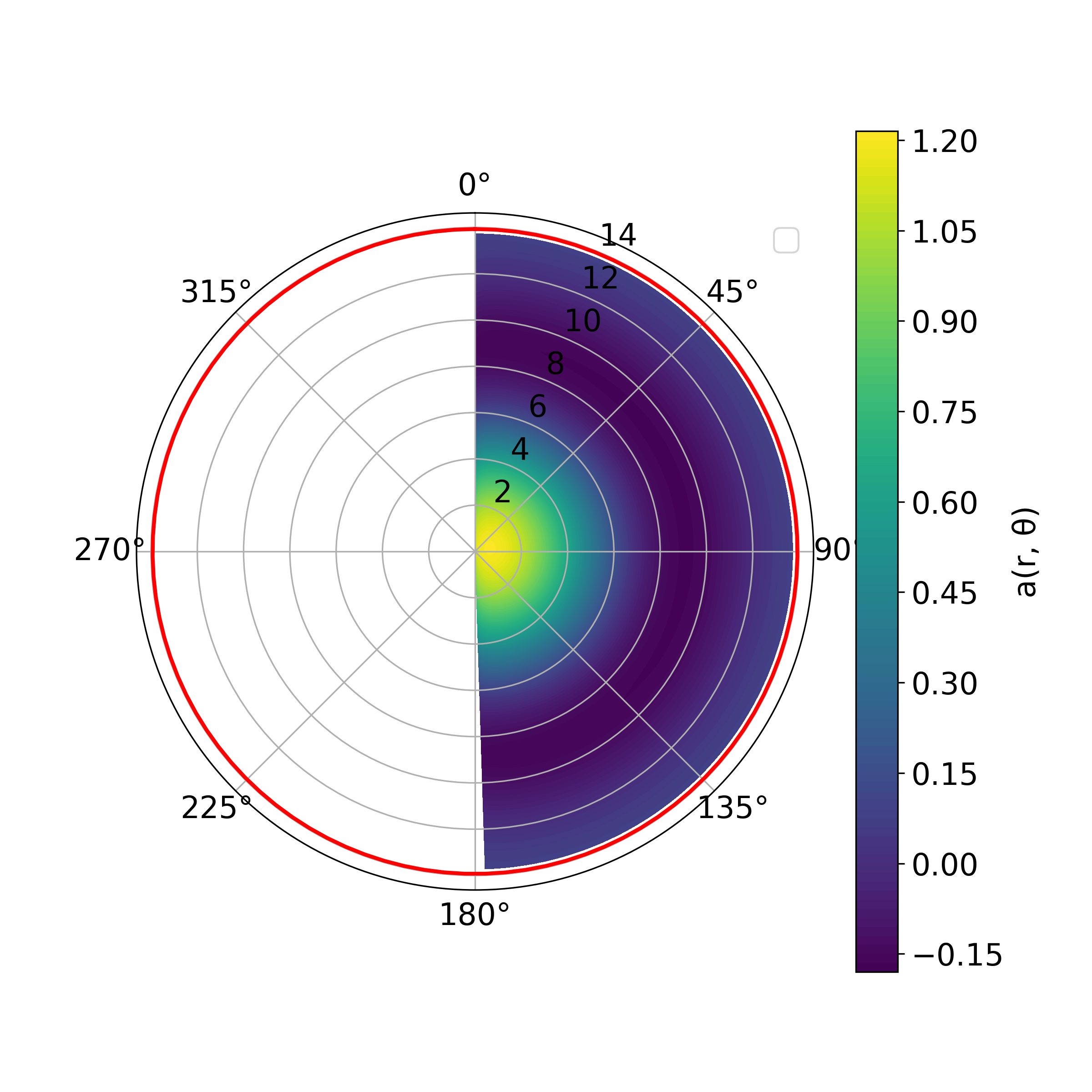}
    \end{minipage}
    \hspace{0.005\textwidth}
    \begin{minipage}[b]{0.48\textwidth}
        \centering
        \includegraphics[width=\linewidth]{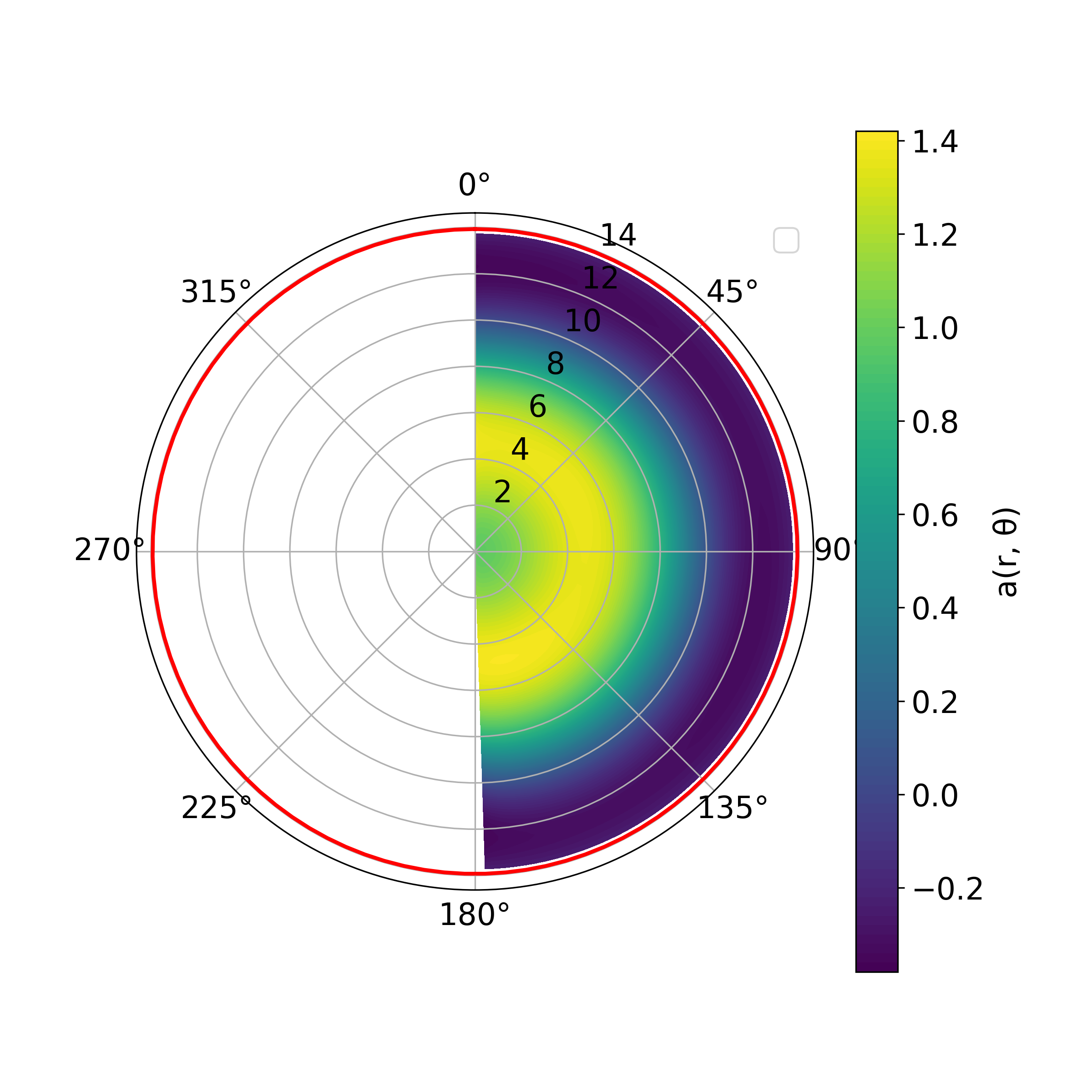}
    \end{minipage}
    \begin{minipage}[b]{0.48\textwidth}
        \centering
        \includegraphics[width=\linewidth]{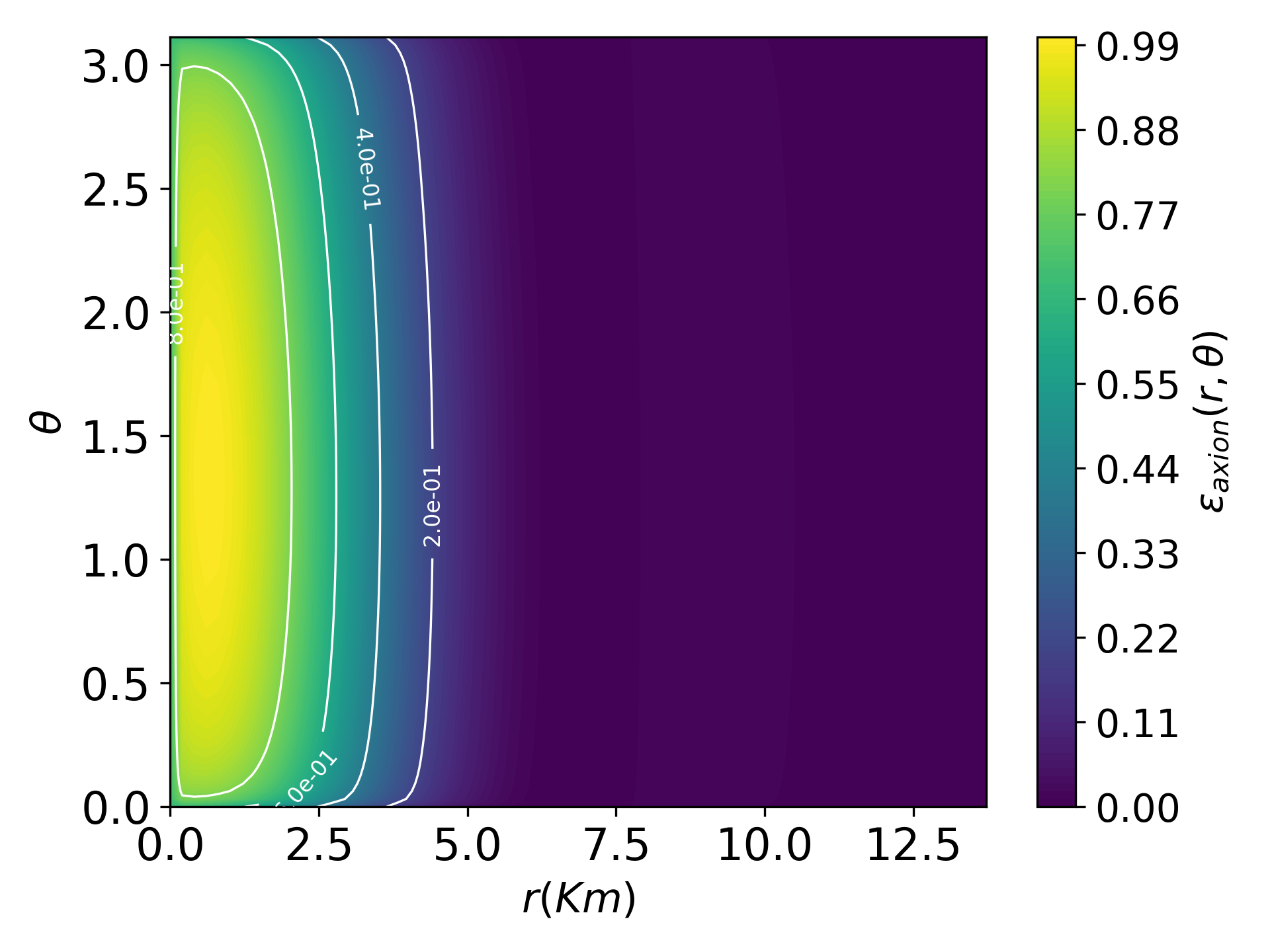}
    \end{minipage}
    \hspace{0.005\textwidth}
    \begin{minipage}[b]{0.48\textwidth}
        \centering
        \includegraphics[width=\linewidth]{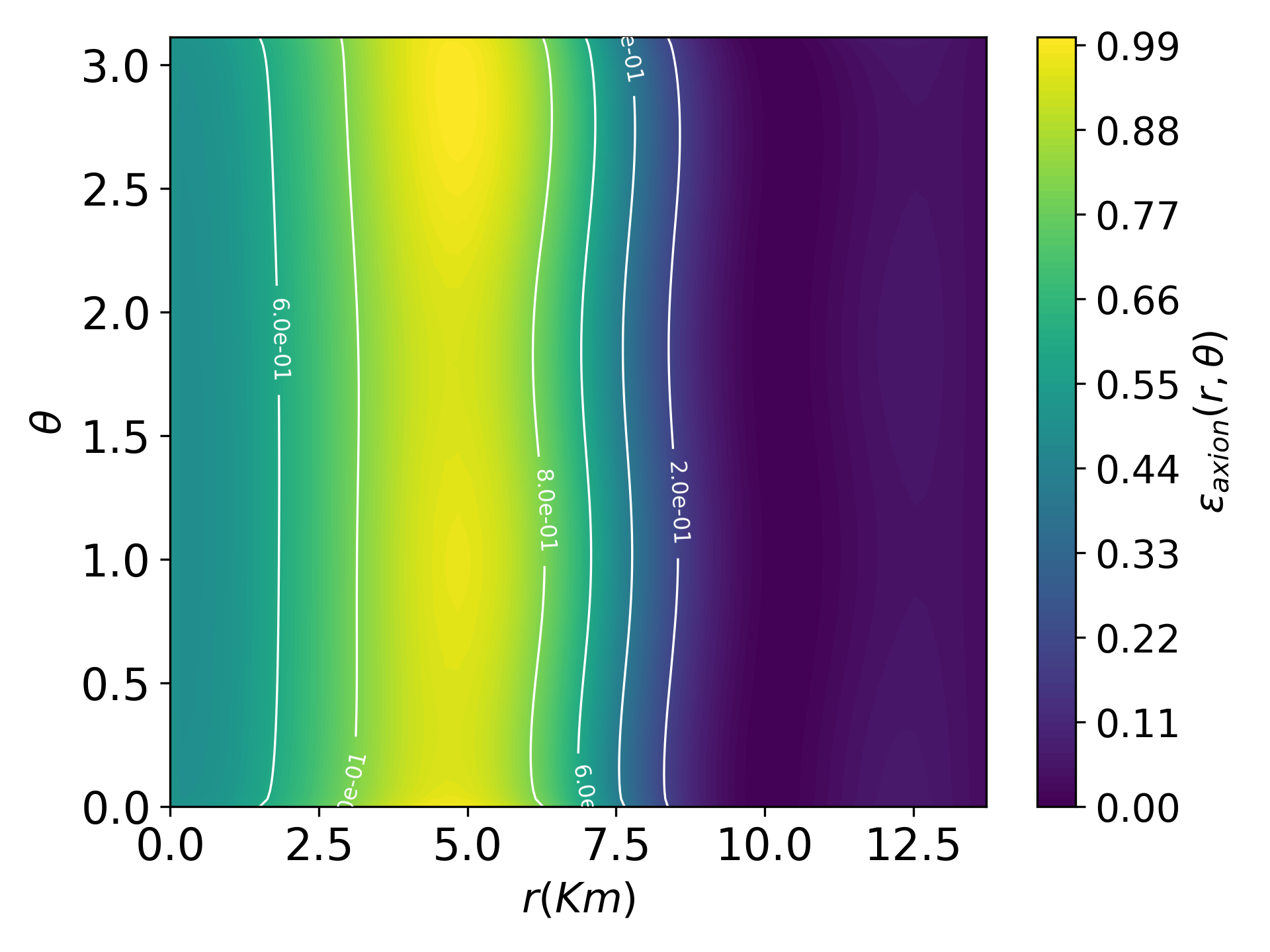}
    \end{minipage}
    \caption{PS-ALP field distribution for a candidate with mass of $m_a=10^{-10}~\text{eV}$ and $g_{a,\gamma}=1.48\times10^{-14} \,\rm GeV^{-1}$ from the source term $G_0$ in Eq. \eqref{axion2d_withS}. The upper panel displays the field distribution in a meridional slice of the star, while the lower panel presents its normalized energy density profile with respect to its value at the origin. The left (right) column corresponds to an average axion field density $\rho_{\text{a}} = 10^{12} \,\rm g/\rm cm^3$ ( $\rho_{\text{a}} =10^{-3} \,\rm g/ \rm cm^3$).  }
    \label{fig:panel_axiones_compacto}
\end{figure}
In Fig. \eqref{fig:panel_axiones_compacto}, we show the axion spatial distribution resulting for $m_a \sim 10^{-10}~\text{eV}$ and $g_{a,\gamma}=1.48\times10^{-14} \,\rm GeV^{-1}$ under the influence of the source term $G_0$ from Eq. \eqref{axion2d_withS}. The upper panel displays the field distribution in a meridional slice of the star, while the lower panel presents the normalized energy density profiles of the axion field with respect to its value at the NS center, $a_0$. In order to explore the axion content effects we consider large versus small fractions, with respect to ordinary matter content. We do not discuss in detail whether this is acquired from an external distribution or has been produced in final state product in stellar reactions, see table I in \cite{vanvlasselaer2024photoproductionaxionsneutrinoscompact}. Accordingly, the left column corresponds to an (averaged over stellar volume) axion field density of $\rho_{\text{a}} \sim  10^{12} \,\rm g/\rm cm^3$, whereas the right column corresponds to a much lower density of $\rho_{\text{a}} = 10^{-3} \,\rm g/ \rm cm^3$. In the latter case, a spatial modulation becomes apparent, and the magnetic field begins to significantly affect the spatial condensation of the axion field. For higher densities at this given axion mass, the $\sim \textbf{E.B}$ source has a negligible influence on the axion field distribution.   
As shown in left panel in Fig. \eqref{fig:panel_axiones_compacto}, the distribution for different values of axion density indicates that for densities comparable to or less than the density of core nuclear matter $\sim 10^{14}\,\rm g/\rm cm^3$, such as in this case of $\sim 0.01\%$ axion fraction they condensate in the center of the star. 

However, as we continue to decrease axion density values, there is a critical point where this factor becomes relevant. This occurs when the axion  mass is on the order of $10^{-21} M_b$, where $M$ is the total  mass of the star or in other words,  at an axion  density $\rho_{a} = 10^{-3} \, \text{g/cm}^3$ (rigth panel in Fig. \eqref{fig:panel_axiones_compacto}). 
In order to see the sensitivity of the axion phase space $(m_a,g_{a,\gamma}, a_0)$ we show in Fig. \eqref{fig:prob} the axion condensate at an average density $\rho_{a}=2.74 \times10^{-19} \rm g\,\,cm^{-3}$. This tiny value in density corresponds to an scenario of weak efficiency in capture/production of axions, although it is enhanced a $\sim10^6$ factor, with respect to local dark matter value $\sim 0.3\,\rm GeV/cm^3$ . The axion mass is set to a lighter value $m_a=10^{-11} \rm \,eV$ with a photon-axion coupling constant of $g_{a,\gamma}=1.48\times10^{-14} \,\rm GeV^{-1}$.}

There is a dramatic difference, and the magnetic field has a noticeable influence at low densities, where the axion field tends to accumulate in certain outer stellar volume regions. This is contrary to the behaviour at higher densities, where the magnetic field has little to no effect, and the axion field becomes entirely concentrated within the inner NS core regions.
It is worth noting the variation of the isocontours of the potential due to the deformation of the star caused by the magnetic field. Not only do the axions begin to accumulate at a certain radius, but also near the poles, although this latter behavior is less noticeable.

\begin{figure}[H]
    \centering
    \begin{subfigure}[b]{0.49\textwidth} 
        \centering
        \includegraphics[width=\textwidth]{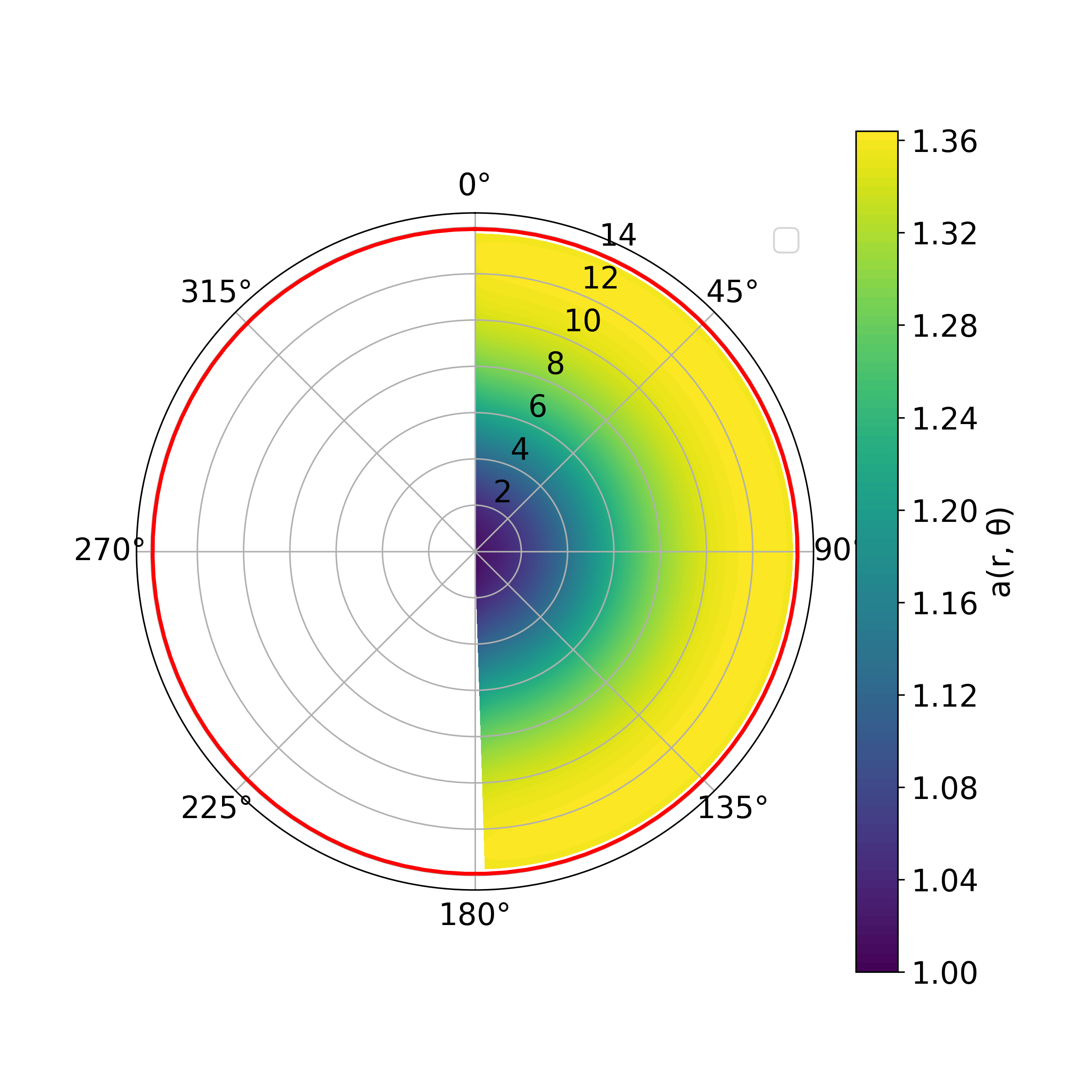} 
        \label{fig:subfig1}
    \end{subfigure}
    \begin{subfigure}[b]{0.49\textwidth}
        \centering
        \raisebox{0.7cm}{\includegraphics[width=\textwidth]{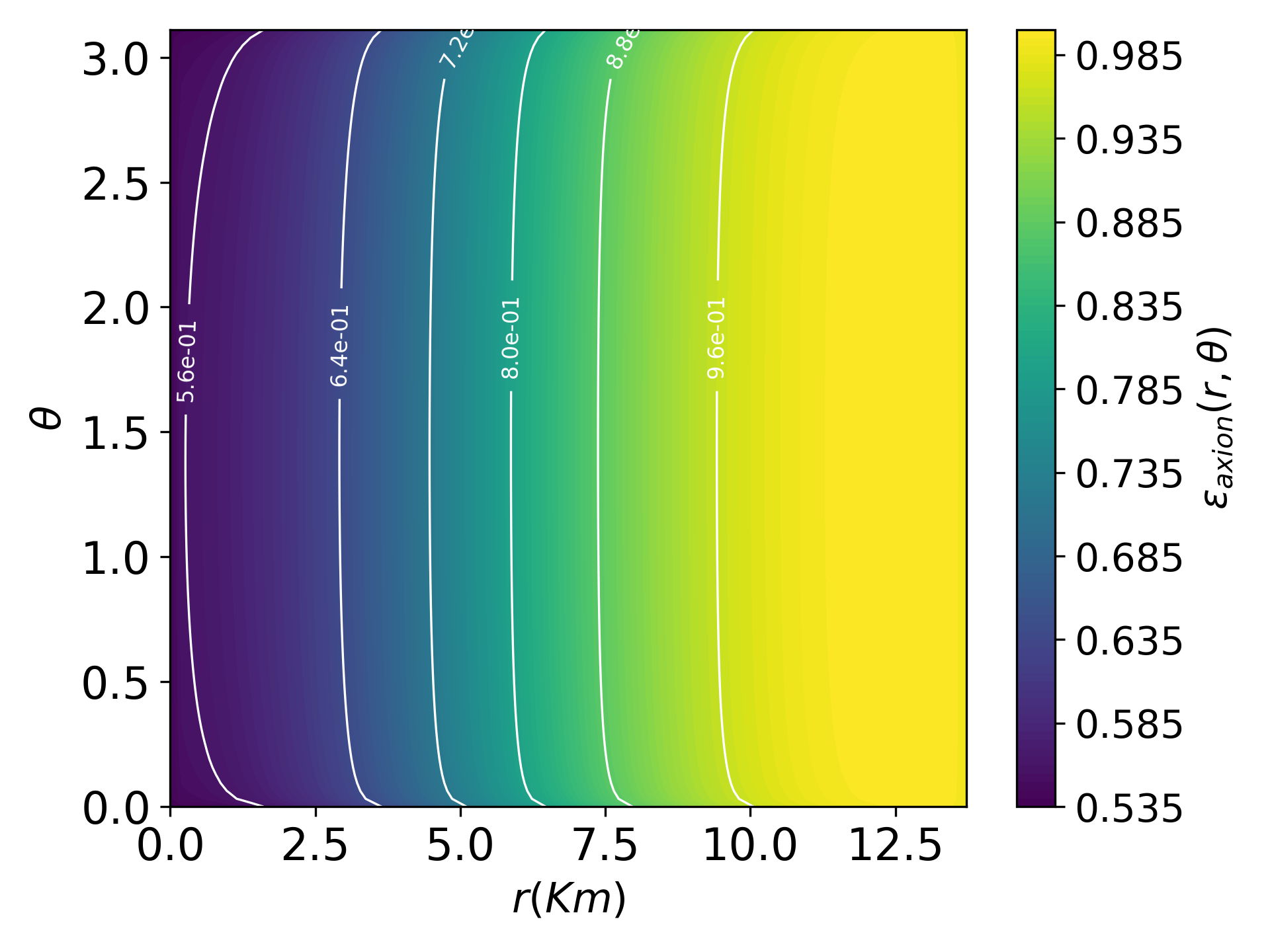}} 
        \label{fig:subfig2}
    \end{subfigure}
    \caption{ PS-Axion distribution for a magnetized NS ($M_{ns}=1.5 M_\odot$ $R=13.93 \rm \, km$) with an axion mass density, $\rho_{a}=2.74 \times10^{-19} \rm g\,\,cm^{-3}$. The axion mass is set to $m_a=10^{-11} \rm \,eV$ with a photon-axion coupling constant of $g_{a,\gamma}=1.48\times10^{-14} \,\rm GeV^{-1}$.}
    \label{fig:prob}
\end{figure}

\subsection{Scalar axion field coupling $Q(F_0,G_0)\equiv F_0 \equiv  g_{\rm eff}(E^2-B^2)$ 
}\label{vi}

We will analyse now the case of  scalar ALPs where the axion field becomes coupled with the Lorentz invariant $F_0$. In this case and assuming the same general expression for the $\bf{E}$ field in Eq. \eqref{E_general} the term $g_{\rm eff}(E^2 - B^2)$ can be written under the form

\begin{align}
g_{\rm eff}(E^2-B^2)&= \frac{e^2 \mu_5^2 \mu_e^2 \Omega^2}{4 c^4 \pi^4 \sigma^2 \hbar^2} + \frac{B^2(r)}{32 c^2 \pi^4 \sigma^2 \hbar^2} \biggl( 8 e^4 \mu_5^2 + \pi^2 \Bigl( c^4 - 32 c^2 \pi^2 \sigma^2 + 32 \pi^2 r^2 \sigma^2 \Omega^2 \Bigr) \hbar^2  \\ \nonumber
& - c^4 \pi^2 \hbar^2 \cos 2\theta \biggr) + \frac{c^2 \sin^2 \theta \left(\partial_r B(r)\right)^2}{16 \pi^2 \sigma^2}  + B(r) \biggl( \frac{e^3 \mu_5^2 \mu_e \Omega}{2 c^3 \pi^4 \sigma^2 \hbar^2} - \frac{c^2 \sin^2 \theta \, \partial_r B(r)}{8 \pi^2 \sigma^2} \biggr)\nonumber
\end{align}

in the rigid rotation approximation with angular velocity ${\bf \Omega}$ aligned to ${\bf B}$ direction in order to determine a convenient expression for $F_0$. In the static case with $\Omega=0$, we find it reduces to  

\begin{align}
g_{\rm eff}(E^2-B^2)=& \frac{1}{32 c^2 \pi^4 \sigma^2 \hbar^2} \biggl( 
    8 e^4 \mu_5^2 + c^2 \pi^2 (c^2 - 32 \pi^2 \sigma^2) \hbar^2- c^4 \pi^2 \hbar^2 \cos 2\theta \biggr)B^2(r) \nonumber \\
& - \frac{c^2 }{8 \pi^2 \sigma^2}B(r)\partial_r B(r)\sin^2 \theta + \frac{c^2 }{16 \pi^2 \sigma^2}\sin^2 \theta \, \left(\partial_r B(r)\right)^2,
\end{align}
or in a compact way
 \begin{equation}
g_{\rm eff}(E^2-B^2)=\beta_1B^2(r) - 2 \beta_2 B(r)\frac{dB(r)}{dr} \sin^2\theta 
    +  \beta_2\left(\frac{dB(r)}{dr}\right)^2\sin^2\theta ,     
\label{eq:reduced_form}
 \end{equation}
with
\begin{align}
&\beta_1\equiv g_{\rm eff}(-1 + \frac{c^2 }{32 \pi^2 \sigma^2} + \frac{e^4 \mu_5^2 }{4  \pi^4 \sigma^2 \hbar^2 c^2} - \frac{c^2\cos2\theta}{32 \pi^2 \sigma^2}),  \,\,\,\,\,\, \beta_2\equiv g_{\rm eff}\frac{c^2}{\sigma^2}. 
\end{align}
In this way we obtain the S-ALP field equation
\begin{align}\label{axion_S_ALP}
    &\frac{1}{f(r,\theta)}\partial_r^2 a(r,\theta)+\frac{1}{r^2k(r,\theta)}\partial_\theta^2 a(r,\theta)+\tilde{\xi_1}(r,\theta)\partial_r a(r,\theta)+ \tilde{\xi_2}(r,\theta)\partial_\theta a(r,\theta) +m^2a(r,\theta) \nonumber \\ &=    \beta_1B^2(r) - 2 \beta_2 B(r)\frac{dB(r)}{dr} \sin^2\theta 
    +  \beta_2\left(\frac{dB(r)}{dr}\right)^2\sin^2\theta. 
\end{align}
Adopting the same numerical values for the parameters as in the PS-ALP case, we can determine that for an axion density of $\rho_{\rm a} = 10^{10} \, \rm g \, \rm cm^{-3}$, i.e., around $0.0001\%$ of the typical central baryonic density, the values are $\beta_1 = -4.26 \times 10^{-25} - 1.21 \times 10^{-52} \rm \cos 2\theta \, \, \rm cm \,\rm erg^{-1}$ and $\beta_2 = 3.83 \times 10^{-50}\, \rm erg^{-1} \,\rm cm^{-1}$. We note that for the dominant $B^2$ term, there is an asymmetry factor depending on $\cos 2\theta$, although it is significantly smaller than the leading term in $\beta_1$. We can derive an expression for this asymmetry by comparing the values of $\beta_1$ along the $z$-direction and the equatorial direction. 
\begin{figure}[H]
    \centering
    \begin{subfigure}[b]{0.49\textwidth} 
        \centering
        \includegraphics[width=\textwidth]{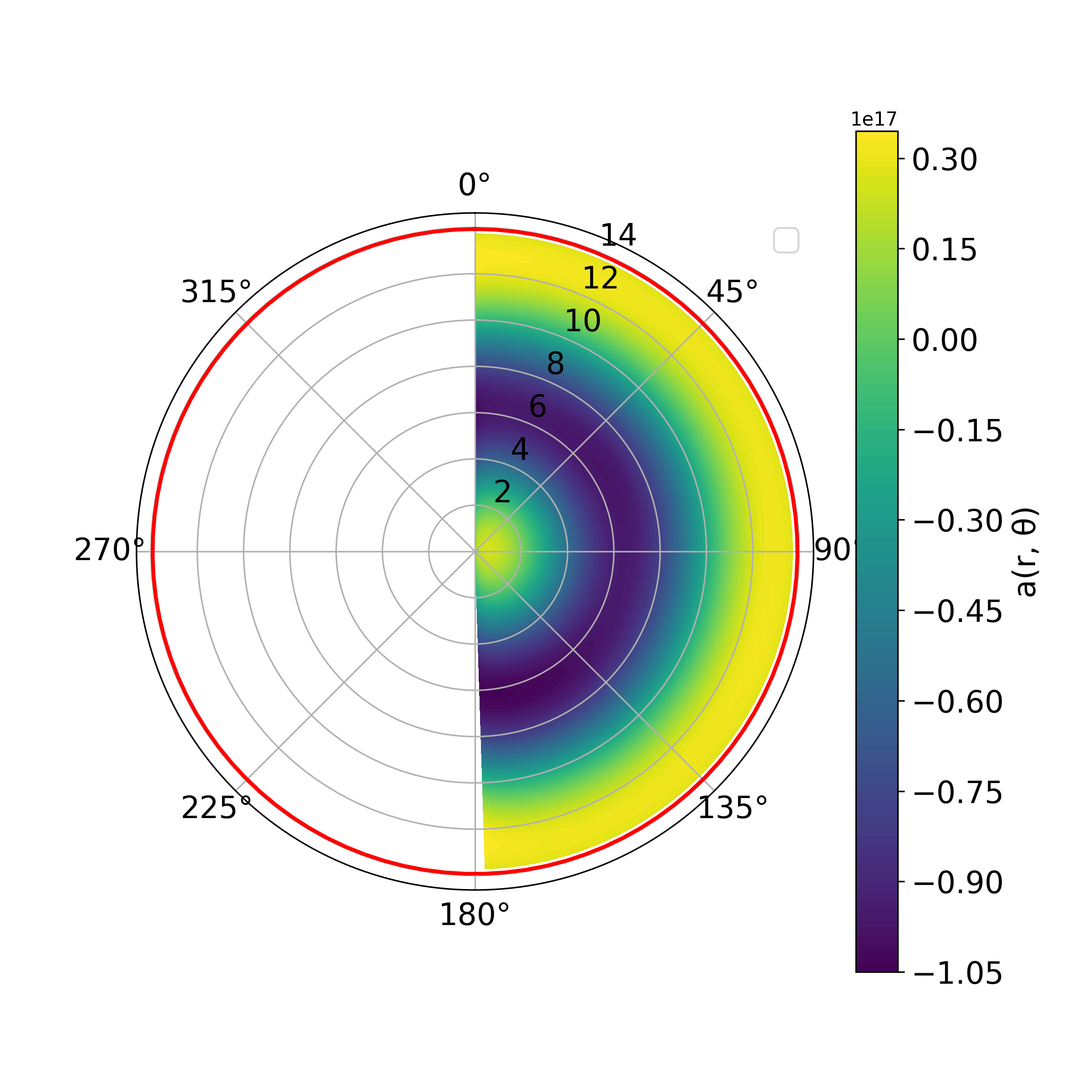} 
        \label{fig:subfig1}
    \end{subfigure}
    \begin{subfigure}[t]{0.49\textwidth}
        \centering
        \raisebox{1.0cm}{\includegraphics[width=\textwidth]{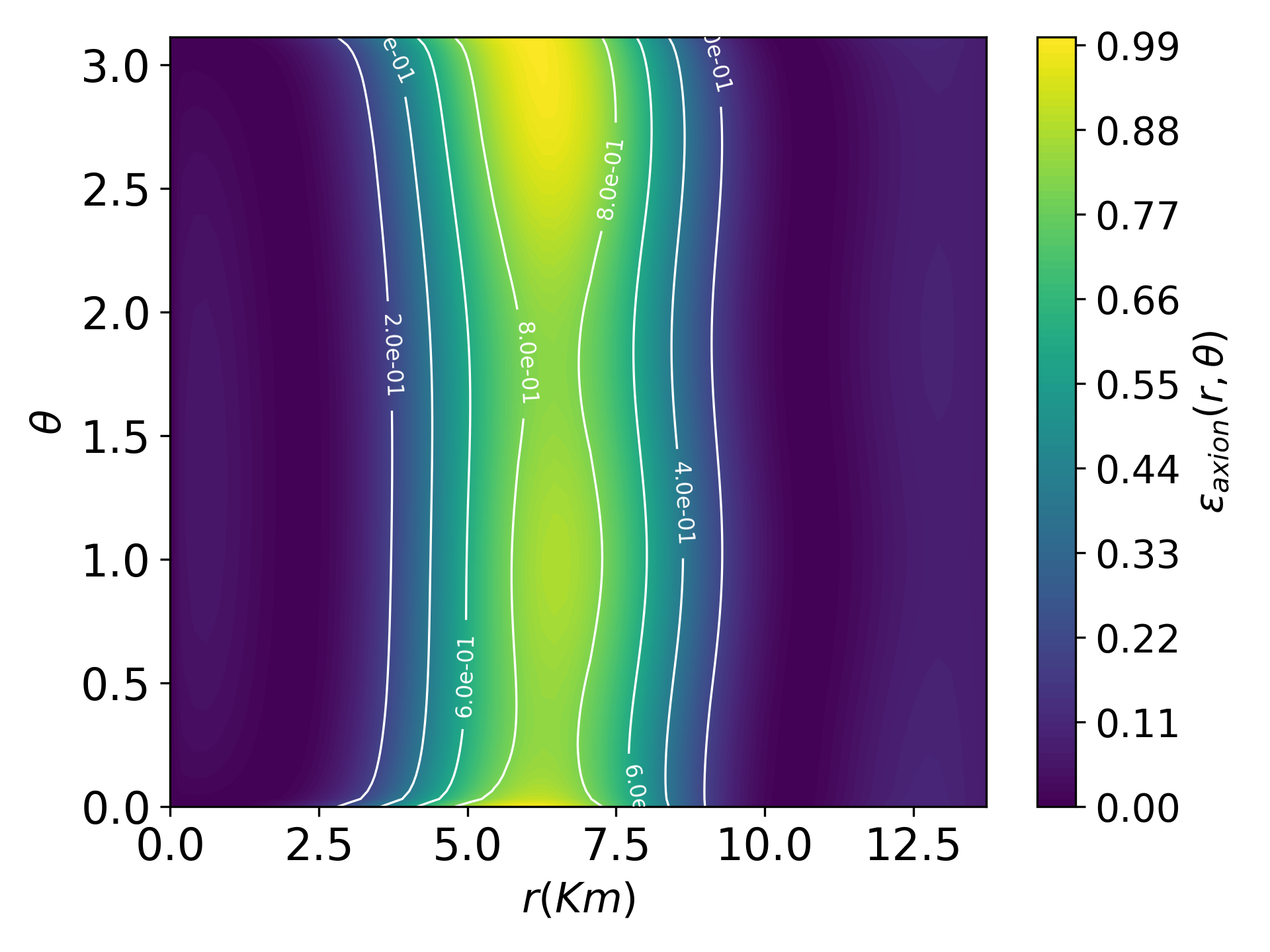}} 
        \label{fig:subfig2}
    \end{subfigure}
    \caption{S-ALP field distribution for a candidate with mass  $m_a=10^{-10} \, \rm eV$ and $g_{a,\gamma}=1.48\times10^{-14} \,\rm GeV^{-1}$ under the influence of a source term $F_0$ from Eq.\eqref{eq:reduced_form}. The left panel shows the axion field in a meridional slice of the star, while the right panel displays the energy density normalized to its central value. In this case, unlike the $G_0$ source, the result is obtained for an axion density $\sim 1\%$ of the typical NS baryonic density, implying that significant effects arise even at low axion  densities }
    \label{fig:S-ALP_graf_1}
\end{figure}
In order to size directionality effects, by subtracting the polar and equatorial directions i.e. $\beta_1[\theta=\frac{\pi}{2}] - \beta_1[\theta=0] = \frac{c^2}{32\pi\sigma^2}$, we obtain that the source has an intrinsic asymmetry that influences the axion distribution. This asymmetry scales with the inverse square of the conductivity so, as expected, a lower conductivity enhances the effect. In the same way, we can evaluate a polar-equatorial asymmetry term for the invariant $F_0$
\begin{align}
F_0[\theta=\frac{\pi}{2}]-F_0[\theta=0]=S_{\rm asym}=\frac{c^2 \left( B(r) - \frac{d B(r)}{d r} \right)^2}{16 \pi^2 \sigma^2}.
\end{align}
From this we see that for a strong magnetic field with weak radial gradient, the source's azimuthal asymmetry will be large, and as a consequence, axion production may become concentrated in specific regions, such as the magnetic poles. This angular dependence, combined with the central axion field amplitude, $a_0$, can locally boost decay/emission channels. The axion distribution for the case with the source term in Eq.\eqref{eq:reduced_form} is obtained by solving  Eq.\eqref{axion_S_ALP} and is shown in Fig.\eqref{fig:S-ALP_graf_1}. We set an axion mass  $m_a=10^{-10} \, \rm eV$ and $g_{a,\gamma}=1.48\times10^{-14} \,\rm GeV^{-1}$.

 On general grounds the energy of massive ALPs in the described scenarios depends on the assumed electromagnetic field modelling and cooling strengths, adding to the gravitational effects described in the assumed perturbed metric. Since a detailed energetic analysis is out of the scope of this work we roughly estimate that the non-relativistic ALP must have an energy  $E_a\sim E_{\mathrm{grav}}+E_{\text {mag }} \simeq-\frac{G M m_a}{R}+g_{a \gamma \gamma}^2 \frac{B^2}{m_a}$. We have neglected in previous the electric field contribution as it is always much smaller than the magnetic field, being only a quadratic correction. Typically for the coupling, there is a threshold value where $E_{\text {mag }} \sim\left|E_{\text {grav }}\right|$. Numerically,  

\begin{equation}
\begin{aligned}
E_{\mathrm{a}} \simeq\;& 
-1.58 \times 10^{-12}\,\mathrm{eV}\,
\left(\frac{M}{1.5\,M_{\odot}}\right)
\left(\frac{13.93\,\mathrm{km}}{R}\right)
\left(\frac{m_a}{10^{-11}\,\mathrm{eV}}\right) \\[6pt]
&+\,3.8 \times 10^{-13}\,\mathrm{eV}\,
\left(\frac{B}{10^{15}\,\mathrm{G}}\right)^{2}
\left(\frac{10^{-11}\,\mathrm{eV}}{m_a}\right)
\left(\frac{g_{a\gamma\gamma}}{10^{-16}\,\mathrm{GeV}^{-1}}\right)^{2},
\end{aligned}
\end{equation}
and the threshold happens when
\begin{equation}
g_{\mathrm{thr}} \simeq 2.04 \times 10^{-16} \mathrm{GeV}^{-1}\left(\frac{M}{1.5 M_{\odot}}\right)^{1 / 2}\left(\frac{13.93 \rm \,km}{R}\right)^{1 / 2}\left(\frac{m_a}{10^{-11} \mathrm{eV}}\right)\left(\frac{10^{15} \mathrm{G}}{B}\right)
\end{equation}
and $g_{a \gamma \gamma} \ll g_{\mathrm{thr}}$ gravitational energy dominates, while for $g_{a \gamma \gamma} \gtrsim g_{\mathrm{thr}}$, magnetic effects rule.

\section{ALP condensates and detectability}
\label{conversion}

As obtained in our setting, axion-photon coupling within the magnetized NS allows a radial and directional dependent static axion condensate. Remarkably, as depicted in Fig.~\eqref{fig:prob}, for certain values of the PS-ALP parameters, here we set $m_a=10^{-11} \rm \,eV$ and $g_{a,\gamma}=1.48\times10^{-14} \rm GeV^{-1}$, axion densities tend to accumulate at intermediate or outer layers, potentially forming a condensate with axion number density, $n_a$, such that  $n_a \lambda_{\text{dB}}^3 \gtrsim 1$, being $\lambda_{\text{dB}}$ the de Broglie associated axion wavelength.  This behavior becomes relevant when considering axion induced radiation emission, as the large number of axions can compensate for the typically low conversion probability.

The probability of axion-photon conversion in an ultradense magnetized and charged particle environment is highly sensitive to the mass difference between the axion ($m_a$) and the effective mass of the photon ($m_\gamma$) within the plasma \cite{pshirkov2009conversion}. This conversion probability reaches a maximum under resonant conditions, 
facilitating efficient mixing between axions and photons. We label $P_{a\rightarrow \gamma}$ as the probability of axion-photon conversion in presence of a magnetic field ${\bf B}$. We assume the axion posses a total energy E and is stabilized in a region with spatial range $L\sim \Delta R<R$  few kilometers wide in stellar radius scale. For the conditions regarding our system in presence of CME discussed in subsection \eqref{v} we obtain

\begin{equation}
    P_{a \rightarrow \gamma}=\left(\frac{g_{a \gamma} B}{\Delta_{\mathrm{q}}}\right)^2 \rm sin ^2\left(\frac{\Delta_{\mathrm{q}} L}{2}\right),
    \label{cmeres}
\end{equation}
 
The general oscillation parameter $\Delta_q$ encapsulates the relevant physics governing the effect. In our previously discussed cases 
\begin{equation}
\Delta_{\mathrm{q}}=\sqrt{\left(\Delta_{\mathrm{a}}-\Delta_{\mathrm{p}}+\Delta_5\right)^2+4 \Delta_{a \gamma}^2},
\end{equation}

and $\Delta_{a \gamma}=\frac{1}{2} g_{a \gamma} B$, $\Delta_{\mathrm{a}}=\frac{m_a^2}{2 E}$ and $\Delta_5=\frac{-\alpha}{2\pi}\mu_5$, the latter being the chiral magnetic effect correction to the plasma contribution $\Delta_{\mathrm{p}}=\frac{\omega_{\mathrm{p}}^{\mathrm{2}}}{2 E}$ and  $\omega_{\mathrm{p}}=\sqrt{\frac{4 \pi \alpha n_e}{m_e}}$ the plasma frequency from electrons with mass and number density, $m_e,n_e$ respectively.  

In our calculation we have obtained that an axion fraction may be confined in equilibrium in the  potential well displaying a non-homogenous distribution inside the stellar volume where they are bound. 
On more general grounds, in the non-relativistic treatment of axions, as candidates to dark matter, their energy in the before mentioned equilibrium configuration depends on the gravitational redshift of the potential well, $z(r)$, provided by the background spherical symmetry $E(r)=m_a \cdot(1+z(r))=m_a \cdot\left(-g_{t t}(r)\right)^{-1 / 2}$  and differs from the case when produced thermally under different channels i.e. involving nucleons through the nucleon-nucleon  (NN) axion bremsstrahlung process $N+N \rightarrow N+N+a$, pion-induced reactions $\pi^{-} +p \rightarrow n +a$ among other, see \cite{raffeltPhysRevLett.60.1793,carenzaPhysRevLett.126.071102}.

In the dense interior of a NS, the plasma frequency $\omega_p$ , which determines the effective photon mass is significantly elevated due to the high electron densities, $n_e$. Typical values of $\omega_p$ in such environments are of the order of tens of keV, see  ~\cite{shternin2009plasma}.
On the other hand, it should also be noted that the CME  introduces a correction term to the plasma frequency that nevertheless does not change this picture. 

Given that the axion mass range of interest in this work, for given couplings, is approximately  $m_a\in [ 10^{-11},10^{-8}]$ eV, the substantial disparity between  $m_a$  and  $m_\gamma=\omega_p$ hardly allows any parameter band for non-vanishing conversion probability  where $\mu_5 \lesssim 10^{-7}$ eV for crust densities \cite{dehman2024origin} at $k_BT\gtrsim 0.1$ MeV in the early proto-NS phase. The expression for determining chiral resonance in Eq. \eqref{cmeres} is given by minimizing $\Delta_q$, from the condition $\Delta_{\mathrm{a}}-\Delta_{\mathrm{p}}+\Delta_5=0$ or equivalently

\begin{equation}
\mu_5=\frac{\pi}{2 \alpha} \cdot \frac{m_a^2-m_\gamma^2}{E_k}.
\end{equation}
However, as discussed, in the interior of the NS this resonance is not likely  since electrons permeate matter at high densities $n_e\gtrsim 10^{25} \rm cm^{-3}$ and $\omega_p \gg m_a$. Only in the magnetosphere this could be, in principle, possible \cite{Prabhu:2021zve}.

The NS crust is a complex region and it is precisely there that for very light axions local condensation may appear. As is well known, in the crust region  ordinary matter displays non-homogeneous structures with shapes ranging from droplets, rods, slabs, tubes and bubbles \cite{Horowitz:2005zb}. These phases are immerse in an electronic degenerate sea to preserve electrical charge neutrality.

Protons in the core of a NS are believed form an Abrikosov phase of a Type II superconductor  in which the external magnetic field leads to an array of flux tubes, each carrying a magnetic flux quantum and $q=2e$ being the charge of a proton Cooper pair.  Even though protons constitute only about less that 1/10 of the mass of the NS they play a crucial role in its rich dynamics because the strong magnetic field leads to the formation of an array of flux tubes, see the recent review \cite{Haskell_2018}. Further, in NS crust protons have been shown to move with sufficient freedom inside  pasta phases, so that they can be considered to be Type II superconductors  \cite{Zhang_2021}.

Briefly, these tubes are filament-like regions where electron density is negligible $m_\gamma \sim \sqrt{n_e}\sim 0$, thus locally the axion-photon conversion probability in Eq. \eqref{cmeres} can be optimized, see below. As the NS crust is populated with such a set of flux tubes (that we call vortex here onwards) we can model this in terms of a periodic array in 2D 
\begin{equation}
\mathbf{B}(\mathbf{r})=\sum_{i=1}^N \mathbf{B}_0 \delta^{(2)}\left(\mathbf{r}-\mathbf{r}_i\right).
\end{equation}

On average, in a given surface $A$, the average B-field is $B_{\mathrm{avg}}=\frac{1}{A} \int_A B(\vec{r}) d^2 r$ 
so that there is a surface density $n_{\text {vortex }}=N/A$ possibly providing a  conversion $P_{a\rightarrow\gamma}$ as a contribution from the set of flux tubes. Typically in  the NS crust we consider $12<\rm log B_{avg}[G]<15$.

\begin{figure}[H]
    \centering
    \begin{subfigure}[t]{0.49\textwidth} 
        \centering
        \includegraphics[width=\textwidth]{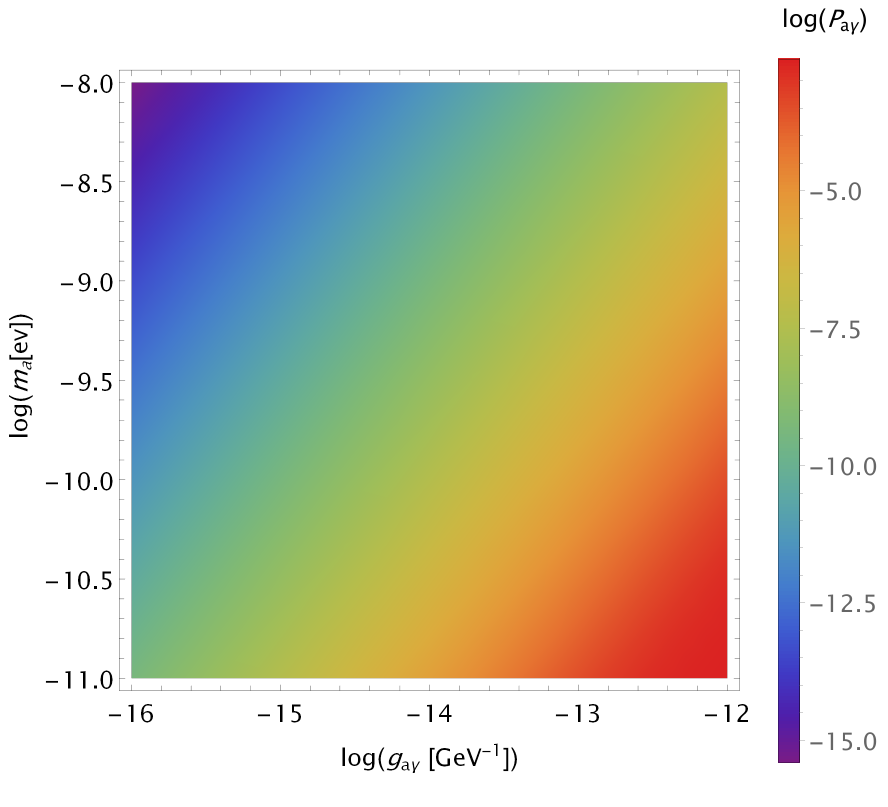} 
        \label{fig:subfig1}
    \end{subfigure}
    \begin{subfigure}[t]{0.49\textwidth}
    \centering
    {\includegraphics[width=\textwidth]{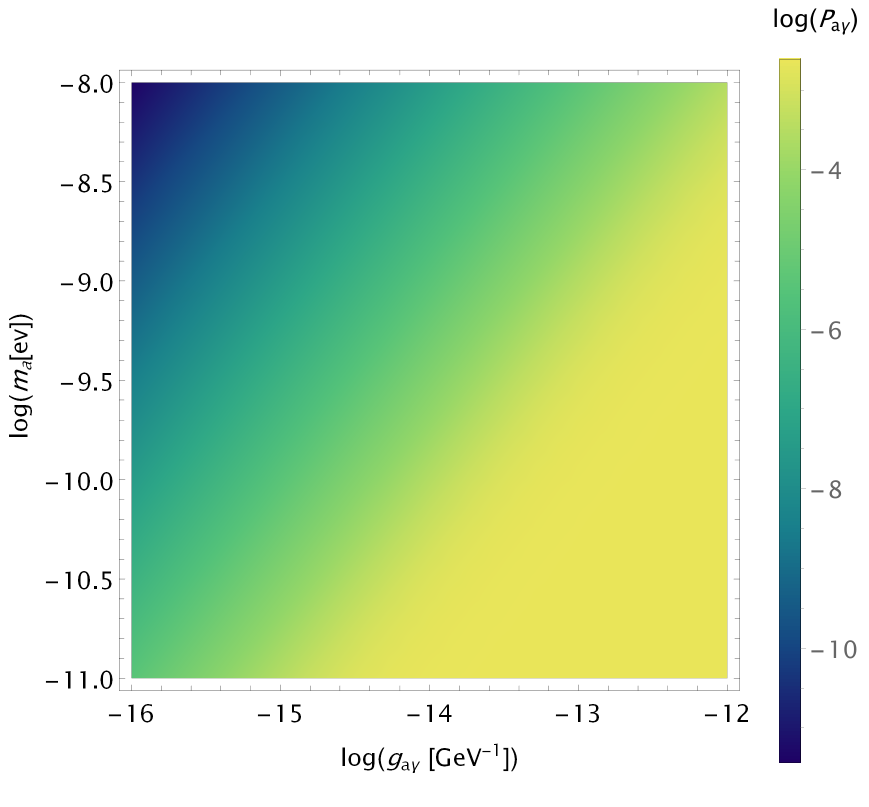}} 
    \label{fig:subfig2}
    \end{subfigure}
    \caption{ALP-photon conversion as obtained from Eq. \eqref{probcoh} for $B_0=10^{13}$ G  (left) and $B_0=10^{15}$ G (right). We assume optimal orientation, see text for details.}
    \label{fig:S-ALP_graf}
\end{figure}

\begin{figure}[H]
    \centering
    \includegraphics[width=0.95\linewidth]{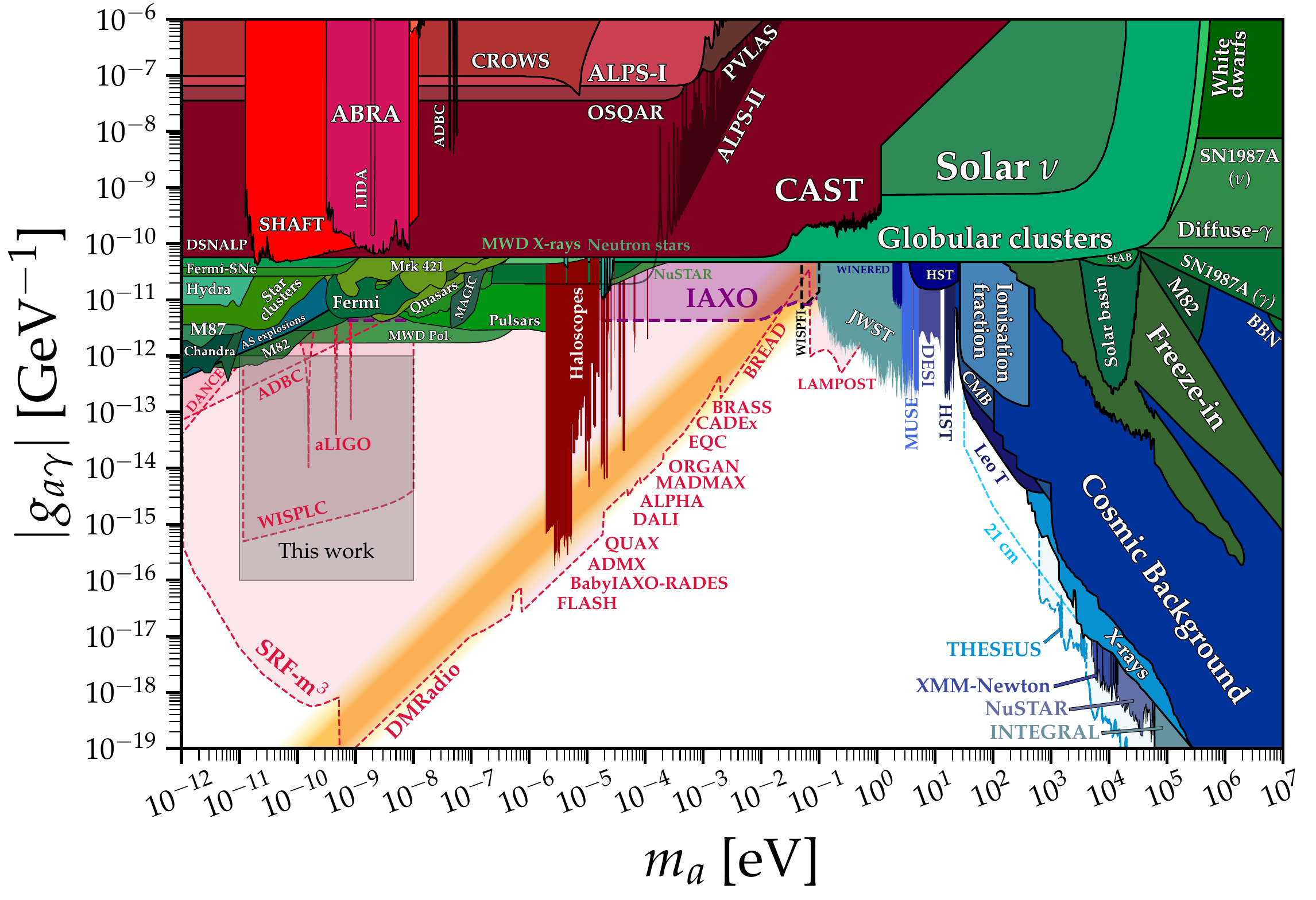}
    \caption{Overview of existing experimental constraints and future projections on the axion–photon coupling  $g_{a \gamma \gamma}$ as a function of the axion mass $m_a$ \cite{AxionLimits}. The gray region labeled \textit{This work} delineates the parameter space relevant for our results, situating them within the context of current and planned axion and axion-like particle searches.}
    \label{fig:placeholder}
\end{figure}

For each flux tube in a type II superconductor, there is a quantized flux unit 
$\Phi_0=\frac{h c}{2 e}\approx 2.07 \times 10^{-7} \mathrm{G} \cdot \mathrm{~cm}^2$ so that the vortex cylindrical section in $\pi \lambda_{\text {vortex }}^2$  with $\lambda_{\text {vortex }}=\sqrt{\frac{m_p^*}{\mu_0 n_p e^2}}$ is approximated by the London length scale. The proton effective mass and proton superconductor density are $m_p^*,n_p$, respectively. In the NS crust it is 
estimated  to be $10\,\rm fm \lesssim \lambda_{\text {vortex }}\lesssim 1000$ fm. If we now associate intervortex distance to the ratio of flux tube to average fields as $d_{\text {vortex }} \sim \sqrt{\frac{\Phi_0}{B_{\rm avg}}}$  and compare de intervortex length given by the ratio $d_{vortex}  \sim \frac{1}{\sqrt{n_{\text {vortex }}}}$ we find $d_{vortex}  \sim 10\lambda_{vortex}$. It is worth noting that our estimate does not consider for this ultra-thin layer any metric perturbations as relevant. From Eqs. \eqref{hmk} the magnetic field enhancement in the flux tubes with respect to the background value will negligibly affect structural features. Within the flux-tube size these tubes do not contain electrons and thus the effective photon mass is null and we have 
\begin{equation}
P_{a \rightarrow \gamma}(\lambda_{vortex})=({g_{a \gamma} B_0 \lambda_{vortex}})^2 \cdot \operatorname{sinc}^2\left(\frac{\lambda_{vortex}}{2} \sqrt{\left(\frac{m_a^2}{2 E}\right)^2+\left(g_{a \gamma} B_0\right)^2}\right),
\end{equation}
with $\operatorname{sinc}(x)=\frac{\sin x}{x}$ and $\Delta_q \approx\sqrt{\left(\frac{m_a^2}{2 E}\right)^2+\left(g_{a \gamma} B_0\right)^2}$. Therefore for the axion coherence length $l_a$ defined from $l_a\sim 1/{\Delta_q}$ and since $d_{vortex}\ll l_{a}$  in our scenario, $\Delta_q \lambda_{vortex} \ll 1$.  Thus $\operatorname{sinc}^2\left(\frac{\Delta_q \lambda_{vortex} }{2}\right) \rightarrow 1$.  A coherent effect in the axion probability is possible when the light axion we consider intersects a forest of effective vortex tubes, $N_{\rm eff}$ so quadratic amplitudes add $\sim N_{\rm eff} ^2P_{1 \text { tube }}$ to yield
\begin{equation}\label{eq:probability}
P_{a \rightarrow \gamma}^{\rm coh}\sim N_{\rm eff}^2 P_{a \rightarrow \gamma}(\lambda_{vortex})= N_{\rm eff}^2 \left({g_{a \gamma} B_0 \lambda_{vortex}}\right)^2 .
\end{equation}
The effective  number of effective flux tubes are those that the axion can intersect in the forest, typically , $N_{\rm eff}=l_a/d_{vortex}$ so that
\begin{equation}
N_{\rm eff} \approx9.8\times\rm 10^{15} \left(\frac{\lambda_{\text {vortex }}}{100 \,\mathrm{fm}}\right)^{-1}\left[0.25\left(\frac{m_a}{10^{-11} \mathrm{eV}}\right)^2+\left(0.195 \cdot \frac{g_{a \gamma}}{10^{-14} \mathrm{GeV}^{-1}} \cdot \frac{B_0}{10^{14} \mathrm{G}}\right)^2\right]^{-1 / 2}.
\end{equation}
Note the real number of flux tubes is expected to be much larger, since the axion may not efficiently resolve them and depends  on  local $n_{vortex}$ being approximated as in  \cite{Graber_2015}. Lastly, it is important to emphasize that
the geometry factors from the transverse component of the magnetic field driving the conversion play a key role, so that the geometry and alignment of the flux-tubes field should be incorporated into an averaged efficiency factor  $(B_{0\perp}/B_0)^2\sim \langle sin^2\theta \rangle\ll1$. From the previous considerations now Eq. \ref{eq:probability} reads

\begin{equation}\label{probcoh}
P_{a \rightarrow \gamma}^{\rm coh}\sim 9.7 \times 10^{-37} N^2_{\rm eff}\, \times\left(\frac{g_{a \gamma}}{10^{-14} \mathrm{GeV}^{-1}}\right)^2\left(\frac{B_0}{10^{14} \,\mathrm{G}}\right)^2\left(\frac{\lambda_{\text {vortex }}}{100 \,\mathrm{fm}}\right)^2
\end{equation}

and note that $P_{a \rightarrow \gamma}^{\rm coh}\sim 3\times 10^{-3}$ for reference values. In Fig.~\eqref{fig:S-ALP_graf} we show the coherent probability in  Eq. \eqref{probcoh} for the NS configuration as in case depicted in Fig.~\eqref{fig:prob}. We assume a flux-tube strength  $B_0=10^{13}$ G  (left) and $B_0=10^{15}$ G (right). We assume optimal orientation with axion-tube alignment so that $B_0\equiv B_{0\perp}$.
Regarding observable signatures this scenario involves an indirect mechanism. Since the optical path inside the crust region is  $\tau\gg1$ photons produced will be most likely absorbed into the local dense environment. Thus considering the axion enhanced density in these outer layers 
we foresee a local rate of deposition energy per unit volume unit time that can be  approximated as  $\varepsilon_{a}=\rho_a v_a P_{a \rightarrow \gamma}$.
In this expression all quantities are related to the conversion zone, assuming the photon energy is half that of the axion rest mass.$v_a \lesssim 0.1$ is  typical velocity in ALP in the distribution that we obtain. 
We find that for reference values before mentioned $P_{a \rightarrow \gamma}^{\rm coh}\sim 3\times 10^{-3}$ local enhanced densities $\rho_a\gtrsim 1.111 \times 10^{-7} \mathrm{~g} \mathrm{~cm}^{-3}$ in the range studied in this work, emissivities may well reach $\varepsilon\sim 10^{9}\rm \,erg/cm^3 s$ being thus comparable to those injecting energy from neutrino-pair bremsstrahlung emissivity due to scattering of electrons by atomic nuclei \cite{Ofengeim_2014}. Thus this energy injected at the base of the crust may well influence the NS surface cooling if efficiently conducted to the exterior \cite{Gud1983ApJ...272..286G}. Current and archival data of isolated cooling NSs from observations by Chandra an other X-ray missions have allowed to test  their interior properties, i.e. Cas A \cite{zhao2025verificationcasneutronstar}. A possibly  detailed observational signature is left for further studies.
In Fig.~\eqref{fig:placeholder} we show  current and projected sensitivities in the ALP parameter space. The gray region depicts the region we study in this work as compared to existing bounds from  \cite{OHare_AxionLimits}. We remark striking coincidence of our studied scenario with that of WISPLC\cite{wisplcPhysRevD.106.023003} and Advaced LIGO experimental settings.
 
In addition, it is likely that the axion condensate is capable of modifying certain properties of the medium, such as its dielectric response, effective mass, magnetic structure—alterations that could give rise to secondary phenomena. It remains to be studied its potentially observability through electromagnetic signatures. Hence, the indirect detection of axions could emerge not solely from direct axion–photon conversions, but also from the subtle yet measurable imprints left on the environment by their collective behavior.

\section{Conclusions}
\label{conclusion}
We have analysed the source terms of scalar and pseudoscalar ALP fields in the magnetized Neutron Star scenario  highlighting differences among them. We have found  several critical insights helping to refine the theoretical framework of axion dynamics in astrophysical environments. First, the axion field exhibits a rapid decay within a region smaller than $1 \, \text{km}$ from the stellar center for axion masses exceeding \(10^{-8} \, \text{eV}\) when propagating freely in the stellar gravitational and B-field induced (non-flat) metric background. This behavior underscores the importance of gravitational interactions and damping effects in dense regions. Second, the inclusion of stellar electromagnetic fields $\textbf{E,B}$, could significantly modify the spatial distribution of axions, displacing their concentration away from the center in a way that depends on the axion field density. For simplicity coupling to baryons or leptons has not been considered, leaving that for a future contribution.

The investigation of the source terms, Lorentz invariants   $\sim\mathbf{E} \cdot \mathbf{B} $ and  $\sim \mathbf{E}^2 - \mathbf{B}^2$ , employing the generalized relation  between $\mathbf{E}$ and $(\mathbf{B})$  from chiral Magneto-Hydrodynamics, revealing distinct phenomena. These results were derived using typical Neutron Star values for extreme magnetic fields in the interior i.e. average central values ($B\sim 10^{18} \, \text{G}$) decreasing with stellar density. 

For the $\mathbf{E} \cdot \mathbf{B} $ term, low-density regimes near the dark matter density scale ($\sim 10^{-3} \, \text{g/cm}^3 $) showed a pronounced magnetic influence, redistributing axions to accumulate preferentially around a characteristic radius inside the star. Conversely, the $ \mathbf{E}^2 - \mathbf{B}^2 $ term introduced an asymmetry tied to the medium’s conductivity and the rest of the magnetic field’s magnitude and its spatial derivative.  \\
Our analysis also suggests that for axion masses $m_a\sim 10^{- 11} \, \rm eV$, axions may accumulate and form a condensate near the neutron star crust. This condensation could alter the local medium, now with a hybrid composition, and coherently enhance the conversion into photons if an areal density of superconductor proton flux tubes is present. This effect opens up possible mechanisms for indirect detection through NS cooling behaviour by locally warming the base of the crust, despite the generally low conversion probabilities within the stellar interior when plasma and chiral resonances are considered.

This work constitutes a first step towards the full resolution of the axion stellar distribution problem, as it focuses solely on the static axion equation of motion. A complete treatment would include  solving the dynamically coupled ALP-Maxwell and gravitational field equations self-consistently, alongside accounting for interactions with baryonic/leptonic matter. The latter becomes increasingly critical toward the stellar center, where baryon-axion couplings may dominate due to extreme densities (\(\gtrsim 10^{15} \, \text{g/cm}^3\)). Such couplings were omitted here but could alter axion dynamics and backreact on the presumed stellar structure.  

These findings emphasize the necessity of incorporating both gravitational and electromagnetic interactions, as well as chiral MHD effects, in modeling axion behavior in compact astrophysical objects. Future studies should further explore the interplay between field gradients, conductivity, baryonic couplings, and axion density thresholds to refine predictions for axion-mediated phenomena in magnetized neutron stars.  
\section*{Acknowledgments}

We acknowledge  partial finantial support by Junta de Castilla y León SA101P24, SA091P24,  MICIU project PID2022-137887NB-I00, Gravitational Wave Network (REDONGRA) Strategic Network (RED2024-153735-E) from Agencia Estatal de Investigación del Ministerio de Ciencia, Innovación y Universidades (MICIU/AEI/10.13039/501100011033) and COST Action COSMIC WISPers CA21106, supported by COST (European Cooperation in Science and Technology)”. D. Suárez-Fontanella acknowledges support from a Banco Santander-USAL Fellowship.

\appendix
\section{Derivation of d'alembertian in curved NS interior}
\label{appendix1}

The explicit derivation of $\frac{\partial_\mu \sqrt{-g}}{\sqrt{-g}}g^{\mu\nu}\partial_\nu a$ requires the inclusion of the metric tensor in Eq. \eqref{metric_per} which accounts for the anisotropic effects of the magnetic field on the stellar structure modeled via Legendre polynomial expansion. If we now rewrite its form into a more generic one we obtain 

\begin{equation}\label{metricGeneral}
d s^2=-h(r,\theta)d t^2+f(r,\theta)d r^2+r^2k(r,\theta)\left(d \theta^2+\sin ^2 \theta d \varphi^2\right)
\end{equation}
where now the following functions are defined
\begin{align}
     &h(r,\theta)= -e^{\nu(r)}-2e^{\nu(r)}(h_0(r)+h_2(r) P_2(\cos \theta)), \label{metric_h}\\
& f(r,\theta)=e^{\lambda(r)}+\frac{e^{2\lambda(r)}}{r}(m_0(r)+m_2(r) P_2(\cos \theta)),\label{metric_f}  \\
& k(r,\theta)=1+2 k_2(r) P_2(\cos \theta) \label{metric_k}.
\end{align}
Then 
\begin{equation}
    \sqrt{-g}=\sqrt{hfk^2}r^2\sin{\theta}
\end{equation}
and the D'alembertian of the axion field can be expanded due to the identity 

\begin{equation}\label{identity}
\nabla_\mu \nabla^\mu \phi=\frac{1}{\sqrt{-g}} \partial_\mu\left(\sqrt{-g} g^{\mu \nu} \partial_\nu \phi\right).
\end{equation}

so we can write for $a$ 
\begin{align}\label{delambertian}
    \nabla_\mu\nabla^\mu a(r,\theta)=&\frac{\partial_\mu \sqrt{-g}}{\sqrt{-g}}g^{\mu\nu}\partial_\nu a + \partial_\mu(g^{\mu\nu}\partial_\nu a),
\end{align} 

Let's start with the first term on the right, due to the symmetry of the metric tensor we can rewrite this term as
\begin{align}
    \frac{\partial_\mu \sqrt{-g}}{\sqrt{-g}}g^{\mu\nu}\partial_\nu a&=\frac{\partial_\alpha \sqrt{-g}}{\sqrt{-g}}g^{\alpha\alpha}\partial_\alpha a \nonumber \\=&\frac{g^{\alpha\alpha}}{\sqrt{hfk^2}r^2\sin{\theta}}\partial_\alpha a\partial_\alpha(\sqrt{hfk^2}r^2\sin{\theta})
\end{align}
for the explicit dependencies
\begin{align}\label{firstterm}
    \frac{\partial_\mu \sqrt{-g}}{\sqrt{-g}}g^{\mu\nu}\partial_\nu a=&\frac{g^{rr}\partial_r a}{\sqrt{hfk^2}r^2\sin{\theta}}\partial_r(\sqrt{hfk^2}r^2\sin{\theta}) + \frac{g^{\theta\theta}\partial_\theta a}{\sqrt{hfk^2}r^2\sin{\theta}}\partial_\theta(\sqrt{hfk^2}r^2\sin{\theta})
\end{align}
now we can expand the partial derivatives 
\begin{align}\label{partialr}
   \partial_r(\sqrt{hfk^2}r^2\sin{\theta})= 2 r\sin{\theta}\sqrt{hfk^2}+\frac{r^2\sin{\theta}}{2\sqrt{hfk^2}}(h'f k^2+hf' k^2+ 2hf kk' )
\end{align}
 where $h'\equiv\partial_r h(r,\theta)$, $f'\equiv\partial_r f(r,\theta)$ and $k'\equiv\partial_r k(r,\theta)$.
For the angular derivative we have
\begin{align}\label{partialtheta}
   \partial_\theta(\sqrt{hfk^2}r^2\sin{\theta})= \sqrt{hfk^2}r^2 \cos{\theta}+\frac{r^2\sin{\theta}}{2\sqrt{hfk^2}}(\oth{h}f k^2+h\oth{f} k^2+ 2hf k\oth{k} )
\end{align}
where $\oth{h}\equiv\partial_\theta h(r,\theta)$, $\oth{f}\equiv\partial_\theta f(r,\theta)$ and $\oth{k}\equiv\partial_\theta k(r,\theta)$.
Then substituting Eqs.\eqref{partialr} and \eqref{partialtheta} into Eq. \eqref{firstterm} 
\begin{align}
    \frac{\partial_\mu \sqrt{-g}}{\sqrt{-g}}g^{\mu\nu}\partial_\nu a=&\frac{g^{rr}\partial_r a}{\sqrt{hfk^2}r^2\sin{\theta}}(2 r\sin{\theta}\sqrt{hfk^2}+\frac{r^2\sin{\theta}}{2\sqrt{hfk^2}}(h'f k^2+hf' k^2+ 2hf kk' )) \nonumber \\
    &+ \frac{g^{\theta\theta}\partial_\theta a}{\sqrt{hfk^2}r^2\sin{\theta}}(\sqrt{hfk^2}r^2 \cos{\theta}+\frac{r^2\sin{\theta}}{2\sqrt{hfk^2}}(\oth{h}f k^2+h\oth{f} k^2+ 2hf k\oth{k} ))
\end{align}
we can reduce it taking as common factor $\sqrt{hfk^2}r^2\sin{\theta}$ and substituting $g^{rr}$ and $g^{\theta\theta}$

\begin{align}
    \frac{\partial_\mu \sqrt{-g}}{\sqrt{-g}}g^{\mu\nu}\partial_\nu a=&(\frac{2}{r}+\frac{h'}{2h}+\frac{f'}{2f}+ \frac{k'}{2k} )\frac{\partial_r a}{f} + (\cot{\theta}+\frac{\oth{h}}{2h}+\frac{\oth{f}}{2f}+\frac{\oth{k}}{k})\frac{\partial_\theta a}{r^2 k}.
\end{align}
Defining
\begin{align}
    &\xi_1(r,\theta)\equiv\frac{2}{r}+\frac{h'}{2h}+\frac{f'}{2f}+ \frac{k'}{k},\\
    &\xi_2(r,\theta)\equiv\cot{\theta}+\frac{\oth{h}}{2h}+\frac{\oth{f}}{2f}+\frac{\oth{k}}{k},
\end{align}
we  have
\begin{align}\label{explicitfirstterm}
    \frac{\partial_\mu \sqrt{-g}}{\sqrt{-g}}g^{\mu\nu}\partial_\nu a=&\frac{\xi_1}{f}\partial_r a + \frac{\xi_2}{r^2 k}\partial_\theta a.
\end{align}

 Let's now focus on the second term on the right in Eq.\eqref{delambertian} i.e
$\partial_\mu(g^{\mu\nu}\partial_\nu a)$. Due to the symmetry of ths metric tensor Eq.\eqref{metricGeneral} and the explicit dependencies of $a$ we can write:
\begin{align}\label{explicitsecondterm}
    \partial_\mu(g^{\mu\nu}\partial_\nu a)&=\partial_\alpha(g^{\alpha\alpha}\partial_\alpha a)\\
    &=\partial_r(g^{rr}\partial_r a)+\partial_\theta(g^{\theta\theta}\partial_\theta a)\\
    &=\partial_rg^{rr}\partial_r a+\partial_\theta g^{\theta\theta}\partial_\theta a +g^{rr}\partial_r^2 a+g^{\theta\theta}\partial_\theta^2 a\\
    &=\frac{1}{f}\partial_r^2 a+\frac{1}{r^2k}\partial_\theta^2 a-\frac{f'}{f^2}\partial_r a-\frac{\oth{k}}{r^2k}\partial_\theta a
\end{align}
so having Eq. \eqref{explicitfirstterm} and Eq.\eqref{explicitsecondterm} now we can rewrite the d'Alembertian as
\begin{align}
    \nabla_\mu\nabla^\mu a&=\frac{1}{f}\partial_r^2 a+\frac{1}{r^2k}\partial_\theta^2 a-\frac{f'}{f^2}\partial_r a-\frac{\oth{k}}{r^2k}\partial_\theta a +\frac{\xi_1}{f}\partial_r a + \frac{\xi_2}{r^2 k}\partial_\theta a \\
    =&\frac{1}{f}\partial_r^2 a+\frac{1}{r^2k}\partial_\theta^2 a+(\frac{\xi_1}{f}-\frac{f'}{f^2})\partial_r a+ (\frac{\xi_2}{r^2 k}-\frac{\oth{k}}{r^2k})\partial_\theta a  
\end{align}
in where defining 
\begin{align}
&\tilde{\xi}_1(r,\theta)\equiv\frac{\xi_1(r,\theta)}{f(r,\theta)}-\frac{f'(r,\theta)}{f(r,\theta)^2}\\
&\tilde{\xi}_2(r,\theta)\equiv\frac{\xi_2(r,\theta)}{r^2 k(r,\theta)}-\frac{\oth{k}(r,\theta)}{r^2k(r,\theta)}
\end{align}
we will have:
\begin{align}
    \nabla_\mu\nabla^\mu a&=\frac{1}{f}\partial_r^2 a+\frac{1}{r^2k}\partial_\theta^2 a+\tilde{\xi_1}\partial_r a+ \tilde{\xi_2}\partial_\theta a  
\end{align}
according with the equation presented in Eq.\eqref{axion2d}

\bibliography{ref_axion}
\bibliographystyle{ieeetr}

\end{document}